\documentclass[twocolumn,showpacs,nofootinbib]{revtex4}
\usepackage{epsf}
\usepackage{graphicx}
\usepackage{dcolumn}

\def\lsim{~\rlap{$<$}{\lower 1.0ex\hbox{$\sim$}}}
\def\gsim{~\rlap{$>$}{\lower 1.0ex\hbox{$\sim$}}}

\newcommand{\BE}{\begin{equation}}
\newcommand{\EE}{\end{equation}}
\newcommand{\BA}{\begin{eqnarray}}
\newcommand{\EA}{\end{eqnarray}}

\def\be{\begin{equation}}
\def\ee{\end{equation}}
\def\bea{\begin{eqnarray}}
\def\eea{\end{eqnarray}}

\def\fun#1#2{\lower3.6pt\vbox{\baselineskip0pt\lineskip.9pt
        \ialign{$\mathsurround=0pt#1\hfill##\hfil$\crcr#2\crcr\sim\crcr}}}

\def\apj{ApJ\,}                 
\def\apjl{ApJL\,}                
\def\apjs{ApJS\,}               
\def\mnras{MNRAS\,}             
\def\aap{A\&A\,}                
\def\aj{AJ\,}                   
\def\physrep{Phys.~Rep.\,}   

\begin{document}
\input epsf
\renewcommand{\topfraction}{0.8}
\preprint{astro-ph/yymmnnn, \today}


\vspace{1cm}

\title{\bf\LARGE Cosmological Information in Weak Lensing Peaks}

\author{\bf Xiuyuan Yang$^{1,5,6}$, Jan M. Kratochvil$^{2}$, Sheng Wang$^{3}$, Eugene A. Lim$^{6,7}$, Zolt\'an Haiman$^{4,6}$, Morgan May$^{5}$}

\affiliation{ {$^1$ Department of Physics, Columbia University, New York, NY 10027, USA}} 
\affiliation{ {$^2$ Department of Physics, University of Miami, Coral Gables, FL 33146, USA}}
\affiliation{ {$^3$ Kavli Institute for Cosmological Physics, University of Chicago, 933 East 56th Street, Chicago, IL 60637, USA}}
\affiliation{ {$^4$ Department of Astronomy and Astrophysics, Columbia University, New York, NY 10027, USA}} 
\affiliation{ {$^5$ Physics Department, Brookhaven National Laboratory, Upton, NY 11973, USA}   }
\affiliation{ {$^6$ Institute for Strings, Cosmology, and Astroparticle Physics (ISCAP), Columbia University, New York, NY 10027, USA}} 
\affiliation{ {$^7$ Department of Applied Mathematics and Theoretical Physics, University of Cambridge, Wilberforce Road, CB3 0WA, UK}} 


{\begin{abstract} 
Recent studies have shown that the number counts of convergence peaks
$N(\kappa)$ in weak lensing (WL) maps, expected from large forthcoming
surveys, can be a useful probe of cosmology.  We follow up on this
finding, and use a suite of WL convergence maps, obtained from
ray-tracing N--body simulations, to study (i) the physical origin of
WL peaks with different heights, and (ii) whether the peaks contain
information beyond the convergence power spectrum $P_\ell$.  In
agreement with earlier work, we find that high peaks (with amplitudes
$\gsim$ 3.5$\sigma$, where $\sigma$ is the r.m.s.  of the
convergence~$\kappa$) are typically dominated by a single massive
halo.  In contrast, medium--height peaks ($\approx0.5-1.5\sigma$)
cannot be attributed to a single collapsed dark matter halo, and are
instead created by the projection of multiple (typically, 4-8) halos
along the line of sight, and by random galaxy shape noise.
Nevertheless, these peaks dominate the sensitivity to the cosmological
parameters $w, \sigma_8$, and $\Omega_m$.  We find that the peak
height distribution and its dependence on cosmology differ
significantly from predictions in a Gaussian random field.  We
directly compute the marginalized errors on $w, \sigma_8$, and
$\Omega_m$ from the $N(\kappa) + P_\ell$ combination, including
redshift tomography with source galaxies at $z_s=1$ and $z_s=2$.  We
find that the $N(\kappa) + P_\ell$ combination has approximately twice
the cosmological sensitivity compared to $P_\ell$ alone.  These
results demonstrate that $N(\kappa)$ contains non-Gaussian information
complementary to the power spectrum.
\end{abstract}}
\pacs{PACS codes: 98.80.-k, 95.36.+x, 98.65.Cw, 95.80.+p}
\maketitle

\section{Introduction}\label{Introduction}

Weak gravitational lensing (WL) by large-scale cosmic structures has
emerged as one of the most promising methods to constrain the
parameters of both dark energy (DE) and dark matter (DM)
(e.g. ref.~\cite{DETF}; see also recent reviews in
refs.~\cite{HJ08,Munshi+08}).  While linear and mildly nonlinear
features in WL maps have been thoroughly explored, an important
question that remains is: how much additional information lies in the
nonlinear features of these maps?  Motivated by this, we recently
investigated a simple nonlinear statistic -- counting peaks in WL maps
directly as a function of their height and angular size
(\cite{PCWLPC}; hereafter Paper I). This statistic does not lend
itself to straightforward mathematical analysis -- it requires
numerical simulations and has received relatively little
attention~(e.g.~\cite{JV00,Wiley+09}) until recent simulation
work~\cite{DH10,PCWLPC}.

In Paper I, we identified peaks in ray-tracing N-body simulations,
defined as local maxima in two-dimensional convergence maps. We found
that the number of peaks as a function of their height $\kappa_{\rm
peak}$ has a sensitivity to a combination of ($w,\sigma_8$)
competitive with other forthcoming cosmological probes.  Dietrich and
Hartlap~\cite{DH10} investigated peak counts as a function of
$\Omega_m$ and $\sigma_8$, and reached qualitatively similar
conclusions.

One result identified in our study is that the cosmological
sensitivity arises primarily from medium--height peaks, with
amplitudes of $\approx 0.5-1.5\sigma$, where $\sigma$ is the r.m.s. of
the WL convergence $\kappa$. Dietrich and Hartlap include only higher
significance peaks in their analysis (with
$\gsim\,2.2\sigma$)\footnote{Note that Dietrich and Hartlap refer to
these as peaks with a signal-to-noise ratio of $\gsim\,3.2$. This is
because our definition of $\sigma$ includes both shape noise and the
cosmological large-scale-structure signal; these are comparable (see
\S~\ref{subsec:mapmaking} below).} but they find a similar trend,
namely that most of the cosmological information is contained in the
lowest significance peaks. In Paper I, we also found that, somewhat
counter-intuitively, the number of the medium-height peaks decreases
with increasing $\sigma_8$."

Motivated by these findings, here we attempt to clarify the physical
origin of the medium amplitude peaks, by identifying collapsed dark
matter halos along sight--lines to individual peaks.  The fact that
the cosmological sensitivity is driven by relatively low--amplitude
peaks raises a potential concern: these peaks may be dominated by
galaxy shape noise, and/or may arise from random projections of
large-scale overdensities in the mildly nonlinear regime.  The counts
of the medium peaks may then offer little information beyond
conventional statistics, such as the power spectrum.  Our second aim
in this paper is therefore to investigate the origin of the
cosmological information content of the WL peaks.  To this end, we
compare peak--height distributions in different cosmologies with those
expected in corresponding Gaussian random fields (GRFs).  Further
improving on Paper I, we are able to provide marginalized constraints
from the combination of peak counts and the power spectrum for the
cosmological parameters $\sigma_8$, $w$, and $\Omega_m$, not just
parameter sensitivity, since we ran a much larger set of simulations.

This paper is organized as follows.
In \S~II, we describe our calculational procedures, including the
creation of the WL maps, the identification of collapsed halos, the
prediction of peak counts in a GRF, and our statistical methodology to
compare maps.
In \S~III, we present our results, which include the matching of peaks
and halos, matching peaks in different cosmologies, and the
comparisons of the simulated peak counts to the Gaussian predictions.
In \S~IV, we offer a detailed discussion of our main results, as well
as of several possible caveats and extensions.
Finally, in \S~V, we summarize our main conclusions and the
implications of this work.

\section{Methodology}
\label{sec:Methodology}

\subsection{Simulating Weak Lensing Maps}
\label{subsec:mapmaking}

We generate a series of 80 cold dark matter N-body simulations for 7
different cosmological models with the code GADGET-2, which include DM
only (no baryons). As our fiducial model, we adopt a $\Lambda$CDM
universe with the following parameters: cosmological constant
$\Omega_\Lambda$ = 0.74, matter density parameter $\Omega_m$ = 0.26,
Hubble constant $H_0=72~{\rm km~s^{-1}Mpc^{-1}}$, dark energy
equation-of-state parameter $w=-1$, and a primordial matter power
spectrum with a spectral index of $n_s=0.96$ and present-day
normalization of $\sigma_8=0.798$.  These values are consistent with
the seven--year results by the {\it WMAP} satellite \citep{WMAP7}.

All simulations use $512^3$ DM particles, in a box with a
size\footnote{Unless stated otherwise, all quantities in this paper
are quoted in comoving units.} of $240h^{-1}$Mpc. This corresponds to
a mass resolution of $7.4\times10^9h^{-1}M_\odot$.  The output of each
simulation run consists of snapshots of particle positions at various
redshifts between $z=0$ and $z=2$, with output redshifts chosen to
span intervals of 80$h^{-1}$Mpc along the line of sight (LOS) in the
fiducial model.\footnote{In cosmologies with different distances, the
same redshift is chosen.}  This interval is shorter, by a factor of
three, than our box size; we truncate the cubes along the LOS to
remove the overlap.  We apply random shifts and rotations to each
snapshot cube, and create gravitational potential planes at each
output by projecting the particle density onto a 2D plane
perpendicular to the line of sight, located at the output redshift,
and solving the Poisson equation. We swap planes from several
independent simulations for the same cosmology in creating the light
cone, to reduce the reuse of the same simulation box and to make the
final WL maps more pseudo-independent. We then follow $2048\times2048$
light rays, starting from $z=0$, and calculate the distortion tensor
and lensing deflection angles at each plane, and produce the final
convergence maps.

\begin{figure}[htp]
\centering
\includegraphics[width=8 cm]{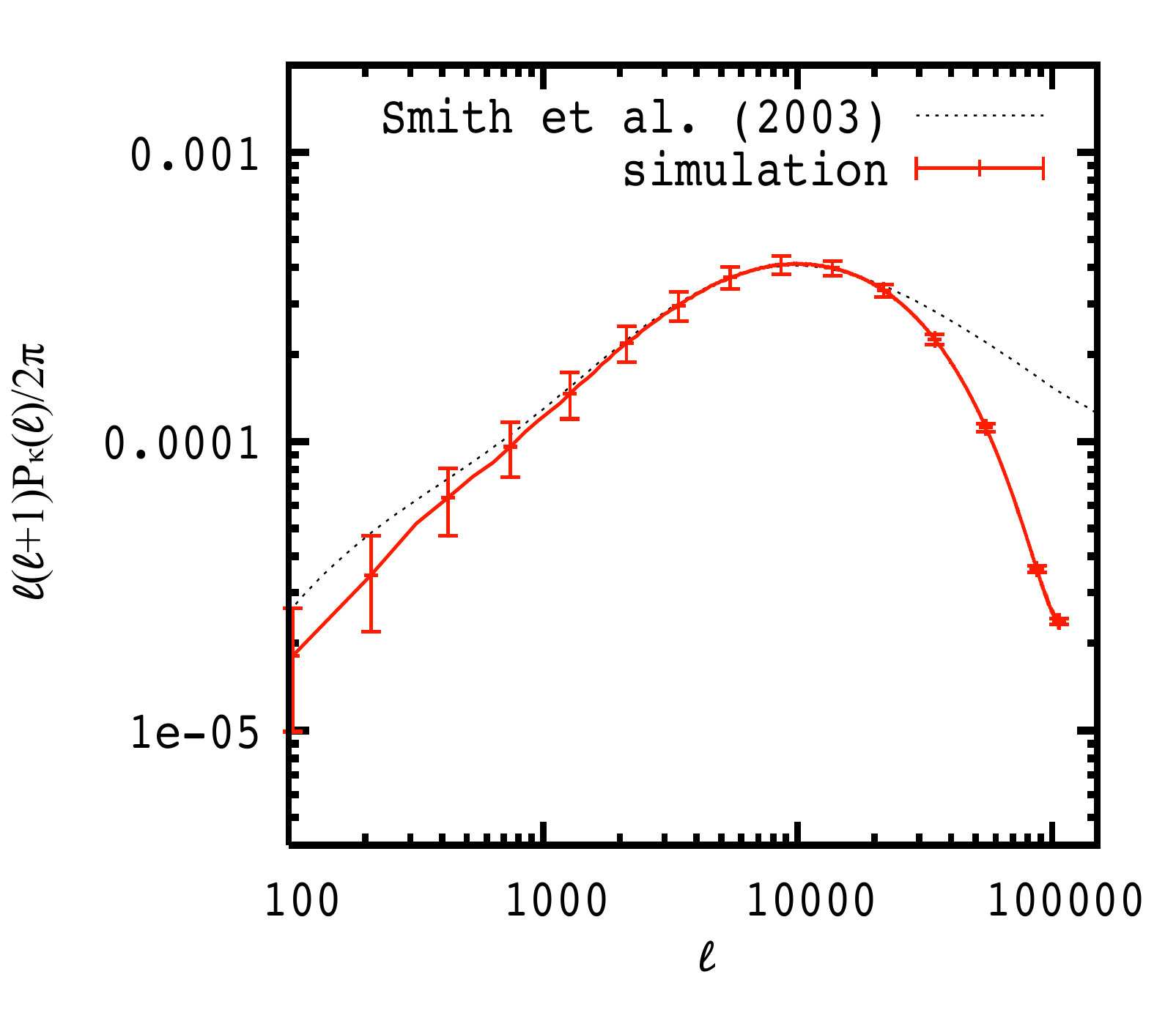}\\ 
\hfill
\caption[]{\textit{Angular power spectrum of the WL convergence as a
    function of spherical harmonic index $\ell$. The solid curve is
    the average over 1000 simulated 12-square-degree convergence maps
    from ray-tracing through 45 independent N-body simulations for the
    fiducial cosmology. The error bars indicate the variance between
    the maps in the bins plotted. The dashed curve is the theoretical
    prediction, based on the 3D nonlinear matter power spectrum
    ~\cite{Smith+03}, with the Limber
    approximation~\cite{Limber1953}. Source galaxies are assumed to be
    located at $z_s=2$. No intrinsic ellipticity noise or smoothing
    were added.}}
\label{fig:WLpowerspec}
\end{figure}

The interested reader is referred to Paper I for more details about
the simulations and the process of making the maps.  We made one
important change, however, which must be high-lighted. While in
Paper~I, the density and potential planes had a resolution of
$2048\times2048$ (same as the resolution used for ray-tracing and the
final convergence map), here we adopted a higher resolution,
$4096\times4096$, for both the density and the potential planes.  This
change has been proven to be important as lower resolution yields a
loss in power at large wave number, as demonstrated
by~\cite{WFWLBSNG}.  In Fig.~\ref{fig:WLpowerspec}, we compare the 2D
angular power spectrum of the WL convergence field at $4096^2$
resolution with the theoretical power spectrum. The latter was
obtained by direct line--of--sight integration, using the Limber
approximation, and the fitting formulae for the nonlinear 3D matter
power spectrum from \cite{Smith+03}, calculated with the code
Nicaea\footnote{Available at
www2.iap.fr/users/kilbinge/nicaea}\cite{Nicaea}.  In the range $400
\lsim \ell \lsim 30,000$, the power spectrum derived from our maps is
not significantly suppressed by either the finite box-size or
resolution, and it agrees well with the theoretical expectation.

Once the maps have been created, we take the redshift-dependent r.m.s.
of the noise in one component of the shear to be~\cite{DCPFCSD}
\BA\label{eq:ellipnoise1}
\sigma_\lambda(z)&=&0.15+0.035z.
\EA 
Note that this corresponds, in the weak lensing limit, to an
r.m.s. ellipticity of $0.30+0.07z$ \citep{Jarvis+03}.

For simplicity, in our analysis we assume that the source galaxies are
located on a source plane at a fixed redshift, with $n_{\rm gal} =
15~{\rm arcmin^{-2}}$. We apply a $\theta_G = 1~{\rm arcmin}$ Gaussian
smoothing to the maps. The noise $\sigma^2_{\rm noise}$ in the
convergence after the Gaussian smoothing then becomes
\BA\label{eq:ellipnoise2}
\sigma^2_{\rm noise}&=&\frac{\langle\sigma^2_\lambda\rangle}{2\pi\theta^2_Gn_{\rm gal}}.
\EA 
For reference, we note that at redshift $z_s=2$, the above gives
$\sigma_{\rm noise}=0.023$, very close to the r.m.s. of the
convergence $\sigma_{\kappa}=0.022$ in the noise-free maps.  The
r.m.s. of the total convergence field, with noise included, is
$\sigma_{\kappa}=0.031$.

\subsection{Suite of Simulated Maps}
\label{subsec:cosmology}

In addition to the fiducial cosmology, we have run N-body simulations
in six other models. Each of these differs from the fiducial model in
a single parameter: we vary $\sigma_8$, $w$, $\Omega_m$ in both
directions, with values $\sigma_8$ = 0.750 and 0.850; $w$ = -0.8 and
-1.2; and $\Omega_m$ = 0.23 and 0.29.  We assume the universe always
stays spatially flat (i.e. $\Omega_\Lambda+\Omega_m=1$).  The seven
different cosmologies will hereafter be referred to as the fiducial,
high-$\sigma_8$, low-$\sigma_8$, high-$w$, low-$w$, high-$\Omega_m$,
and low-$\Omega_m$ models, respectively, as summarized in Table
~\ref{tab:Cosmologies}.  In each of these N-body runs, we create 1000
different WL maps with source galaxies at $z_s=1$, and another 1000
maps with galaxies at $z_s=2$. Each map covers a solid angle of
$3.46\times3.46$ degrees.  All maps were created by mixing potential
planes randomly among five different N-body runs, with independent
realizations of the initial conditions, in the given cosmology.
Finally, we created an additional 1000 control maps, using the planes
from 45 additional independent N-body runs in the fiducial model.
Having 9 times more strictly independent realizations allows us to
compute the covariance matrix more accurately (needed for computing
$\Delta\chi^2$; see below), and to check the robustness of our results
to different realizations of the fiducial model.

\begin{table}
\begin{tabular}{|l|c|c|c|c|} 
\hline
 & $\sigma_8$ & $w$ & $\Omega_m$ & \# of sims \\
\hline
Fiducial & 0.798 & -1.0 & 0.26 & 5\\
Control & 0.798 & -1.0 & 0.26 &45\\
High-$\sigma_8$ & 0.850 & -1.0 & 0.26 & 5\\
Low-$\sigma_8$ & 0.750 & -1.0 & 0.26 & 5\\
High-$w$ & 0.798 & -0.8 & 0.26 & 5\\
Low-$w$ & 0.798 & -1.2 & 0.26 & 5\\
High-$\Omega_m$ & 0.798 & -1.0 & 0.29 & 5\\
Low-$\Omega_m$ & 0.798 & -1.0 & 0.23 & 5\\
\hline
\end{tabular}
\caption[]{\textit{Cosmological parameters varied in each model.  The
universe is always assumed to be spatially flat
($\Omega_\Lambda+\Omega_m=1$).}}\label{tab:Cosmologies}
\end{table}

\subsection{Halo Finding }
\label{subsec:halofinder}

We use the publicly available AMIGA halo finder (\cite{AHF}; hereafter
AHF) to identify collapsed halos in our N-body runs.  AMIGA finds
halos based on an iterative density refinement scheme. Its output
consists of the 3D positions of the halos, and, importantly for us,
the tagged set of particles belonging to each halo.  The virial radius
of a halo is such that when a sphere is placed at the halo's location,
with a radius $r_{\rm vir}$, the overdensity $\bar{\rho}(r_{\rm vir})$
is
\BA
\label{eq:AHFhalo1}
\bar{\rho}(r_{\rm vir})&=&\Delta_{\rm vir}\rho_b, 
\EA
where $\rho_b$ is the background baryon density, and where
$\Delta_{vir} = 180$ is adopted in this study. The mass of the halo is
then simply given by
\BA
\label{eq:AHFhalo2}
M_{\rm vir}&=&4\pi\rho_b\Delta_{\rm vir}r^3_{\rm vir}/3.
\EA 

\begin{figure}[htp]
\centering
\includegraphics[width=8 cm]{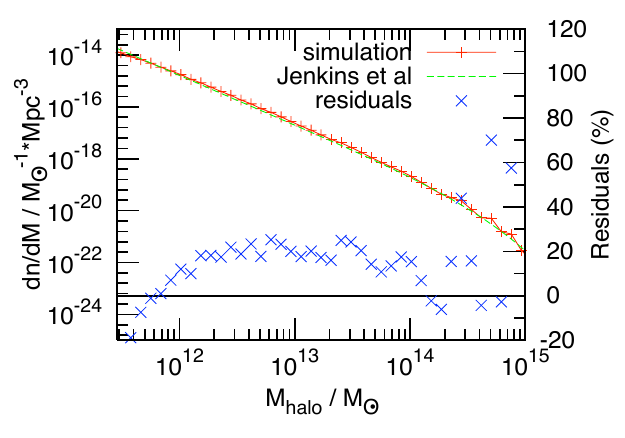}\\ 
\hfill
\caption[]{\textit{Halo mass function, produced with the AMIGA halo
finder \cite{AHF}, in our N-body simulation of the fiducial model at
$z=0$, compared to the fitting formula from Jenkins et
al. \cite{Jenkins+01}.}}\label{fig:massfunction}
\end{figure}

As a simple test of both our N-body simulations and our implementation
of the halo finder, we reproduce the fitting formula (their equation
B3.)  for the halo mass function reported by Jenkins et
al.~\cite{Jenkins+01}.  An example of this comparison is shown, in our
fiducial model at $z=0$, in Fig.~\ref{fig:massfunction}.  Overall, we
find excellent agreement, with an accuracy of 25\% or better up to a
halo masses of $2\times 10^{14} {\rm M_\odot}$. For larger masses,
there is a large scatter.

One of the goals in this study is to identify halos contributing to
each individual convergence peak. To do this, starting from each peak,
we follow the light ray, and record the information (masses and
location) of the halos found within a light cone centered on the peak,
with a radius of 3 arcmin.  This radius is chosen to be three times
the smoothing scale. We have verified that doubling the radius of the
light cone does not change our halo matching results below -- in the
sense that no additional halos are identified that contribute
significantly to the total convergence of a peak (see detailed
discussion below).  To be consistent with the perpendicular projection
of the particle density in each simulation snapshot, the light rays
consist of a series of parallel line segments, which are perpendicular
to the potential planes.  The coordinates where the segments cross the
potential planes are determined by the lensing deflection angle,
computed and stored during the ray-tracing analysis.  For the same
reason, the light cones are composed of a series of parallel cylinders
with a radius of 3 arcmin, centered on the corresponding light ray.

As explained above, we truncate the simulation snapshots, in order to
generate more independent realizations of maps.  As a result,
occasionally, parts of halos that happen to be located near the plane
of the truncation can be unphysically ``cropped''.  These cropped
halos become important only when they are sufficiently massive to
contribute to the convergence of a peak, and when they are located
near (within a fraction of their virial radius) one of the two
truncation planes (either in the front or the back).  Given that halo
virial radii are of order $\sim1h^{-1}$Mpc, and our truncated box size
is 80$h^{-1}$Mpc, the probability that the latter condition is
satisfied is $\sim(1+1)/80\sim 2.5\%$.  We therefore simply restrict
our halo catalog only to halos that do not touch the edges. We have
checked that neglecting the cropped halos does not significantly
affect our results.

Our simulations resolve the density structure of individual halos with
masses of $M\gsim {\rm few}\times 10^{11}~{\rm M_\odot}$.  However,
each map contains on the order of $10^3$ peaks, and more than a dozen
halos can contribute to the total convergence of a single peak;
therefore, computing the exact contribution of each halo to each peak
is computationally impractical.  Instead, we replace each halo by a
spherically symmetric NFW~\cite{UDPHC} halo with the same virial mass
$M_{\rm vir}$.  For a given impact parameter $d$ (defined as the
angular distance between the halo center and the point of closest
approach of the light ray corresponding to the peak), redshift $z$,
and mass $M_{\rm vir}$, the contribution of the off-center halo to the
convergence peak can then be calculated analytically.

More specifically, the density profile is assumed to follow
\BA\label{eq:NFW1}
\rho_{\rm nfw}(r)&=&\frac{\rho_s}{(r/r_s)[1+(r/r_s)]^2},
\EA 
where $r$ is the radius from the halo center, and $r_s$ and $\rho_s$
are a characteristic radius and density. The profile is truncated at
$r_{200}$, inside which the mean overdensity with respect to the
critical density of the universe at redshift $z$ is 200. We adopt the
concentration parameter $c_{\rm nfw}=r_{200}/r_s=5$ in this paper.
The convergence due to the halo, given an extended redshift
distribution of the background galaxies, is then given by
\BA
\label{eq:convergence} \kappa(\phi)&=&\frac{4\pi
G}{c^2}\frac{\Sigma(\phi)\chi_z}{(1+z)}\frac{\int^\infty_z\,dz^\prime(dn/dz^\prime)(1-\chi_z/\chi_{z^\prime})}{n_{\rm
tot}}.  
\EA
Here $\Sigma(\phi)$ is the projected surface density of the halo
(given explicitly in ref.~\cite{TJ03}; see their equations 26-27),
$\chi_z$ is the comoving distance to redshift $z$, $dn/dz$ is the
surface number density of background galaxies per unit redshift, and
$n_{\rm tot}$ is the mean total surface density.  The latter is taken
to be $n_{\rm tot}\delta(z-z_s)$ in this paper, with $z_s$ as the
source redshift. Finally, we use a Gaussian window function to smooth
the convergence induced by the halo,
\BA
\label{eq:smooth1}
\kappa_G&=&\int\,d^2\phi W_G(\phi)\kappa(\mid\vec{\phi}-\vec{\phi_0}\mid)
\EA 
\BA\label{eq:smooth2}
W_G(\phi)&=&\frac{1}{\pi\theta^2_G}\exp(-\frac{\phi^2}{\theta^2_G})
\EA 
\BA\label{eq:smooth3}
\mid\vec{\phi_0}\mid&=&\frac{d(1+z)}{\chi_z}
\EA 
where the center of the smoothing kernel ($\vec{\phi}=0$) is set to
the angular position of the light ray corresponding to a peak, and
$\vec{\phi_0}$ is the angle toward the halo center.  The smoothing
scale $\theta_G$ is chosen to be 1 arcmin, as in the simulated maps.

\begin{figure}[htp]
\centering
\includegraphics[width=8 cm]{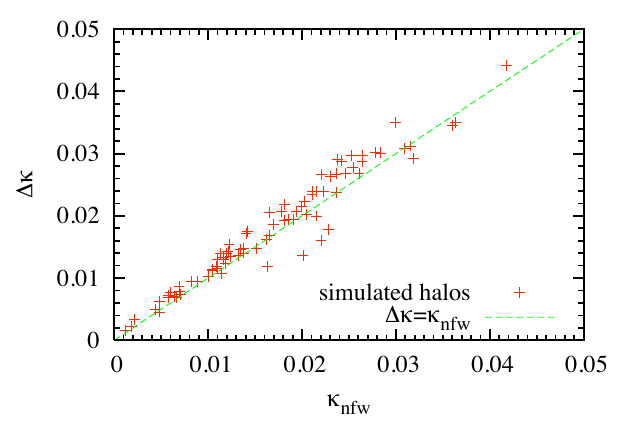}\\ \ \\
\hfill
\caption[]{\textit{Comparison of the convergence $\kappa_{\rm nfw}$,
produced by halos with an NFW profile, and the difference
$-\Delta\kappa$ in the simulated convergence map induced by
artificially removing the halo from the 3D simulation box. 81 halos,
identified in a $z_s=1$ map generated in our fiducial model, were used
for this exercise.  No galaxy noise was added to the maps.  The NFW
assumption works well, with a fractional bias of only -6.7 percent,
and a scatter of 15 percent, relative to $\Delta\kappa$.
}}\label{fig:diff_nfw}
\end{figure}

To check the accuracy of the NFW approximation for the convergence, we
selected $81$ halos with masses in the range $1.5\times 10^{12}~{\rm
M_\odot}< M < 1.5\times 10^{14}~{\rm M_\odot}$ in one of the
realizations of our fiducial model.  For each halo, we record the
value of the convergence $\kappa$ in the map (with sources galaxies at
$z_s=1$), in the pixel located in the direction toward the halo
center.  We then remove the halo from the simulation box, and repeat
the ray-tracing procedure discussed above, to compute a new value
$\kappa_0$ at the same position, but without the halo. In
Fig.~\ref{fig:diff_nfw}, we show the actual difference
$\Delta\kappa\equiv\kappa-\kappa_0$, against the value $\kappa_{\rm
nfw}$ expected based on the NFW halo model.  As the figure shows, the
NFW assumption works accurately, although it results in a slight
underestimate of the convergence.  The points in the figure yield an
average fractional bias of
$\langle(\kappa_{\rm nfw}-\Delta\kappa)/\Delta\kappa\rangle = -0.067$
and an r.m.s. scatter of
$\langle(\kappa_{\rm nfw}-\Delta\kappa)^2/\Delta\kappa^2\rangle^{1/2}=0.15$.

\subsection{Gaussian Random Field Predictions}
\label{subsec:GRF}

One of the key questions to be answered in this paper, is the extent
to which the peak counts contain information beyond traditional
measures, such as the power spectrum.  For example, if the peaks were
produced only by a combination of pure galaxy shape noise (which is
Gaussian by assumption) and linear fluctuations in the matter density,
then their statistics would be fully described by a Gaussian random
field (GRF).  The majority of the high peaks are known to be
associated with collapsed, nonlinear objects, and their statistics
will clearly be non-Gaussian. However, the extent to which
lower-amplitude peaks are non-Gaussian is not clear ab-initio. As
mentioned in the Introduction, these peaks contain most of the
cosmological information, and therefore the departure in the
statistics of these peaks from Gaussian predictions is important to
understand and quantify.

\begin{figure}[htp]
\centering
\includegraphics[width=8 cm]{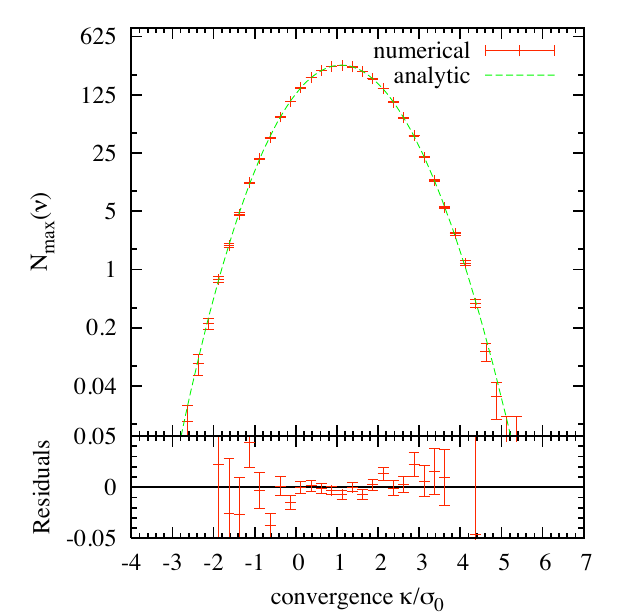}\\ \ \\
\hfill
\caption[]{\textit{Number of convergence peaks in a Gaussian random
    field, as a function of their height measured in units of the
    standard deviation of the convergence, $\sigma_0$.  The data
    points show the number of peaks in bins of width
    $\Delta\kappa=0.25\sigma_0$, obtained by averaging counts in 200
    random Monte Carlo realizations of a 2d GRF.  The input power
    spectrum was calculated from the non-linear matter power spectrum
    in Smith et al.~\cite{Smith+03} in our fiducial cosmology, with
    source redshift $z_s=2$, including galaxy noise and smoothing.
    The map size, after excluding 40 pixels along each edge, is
    $2.88\times2.88~{\rm deg}^2$.}}\label{fig:GRFpeaks}
\end{figure}

As a simple test, we directly compare our simulated peak counts with
those expected in a GRF.  Fortunately, the peak counts in a
two--dimensional GRF, and their distribution in height, are
predictable analytically~\cite{SCBRF}.  In fact, they depend only on
the first and second derivatives of the correlation function on small
scales (or, equivalently, the first two moments of the power
spectrum). For completeness, we reproduce the relevant equations
here. The differential number of maxima per unit solid angle, $n_{\rm
max}(\nu)$, with height in the range $\nu$ to $\nu+d\nu$, where $\nu$
is measured in units of the standard deviation $\sigma_0$ of the
random field, is given by
\BA\label{eq:GRFeq1} 
n_{\rm max}(\nu)d\nu&=&\frac{1}{2\pi\theta^2_*}\exp(-\nu^2/2)\frac{d\nu}{(2\pi)^{1/2}}G(\gamma,\gamma\nu)
\EA
where
\BA
\label{eq:GRFeq2}
G(\gamma, x_*)&=&\nonumber(x^2_* -
\gamma^2)[1-\frac{1}{2}{\rm erfc}\{\frac{x_*}{[2(1-\gamma^2)]^{1/2}}\}]\\
\nonumber&&+x_*(1-\gamma^2)\frac{\exp\{-x_*^2/[2(1-\gamma^2)]\}}{[2\pi(1-\gamma^2)]^{1/2}}\\
\nonumber&&+\frac{-x^2_*/(3-2\gamma^2)}{(3-2\gamma^2)^{1/2}}\\
&&[1-\frac{1}{2}{\rm erfc}\{\frac{x_*}{[2(1-\gamma^2)(3-2\gamma^2)]^{1/2}}\}]
\EA
\BA\label{eq:GRFeq3}
\gamma&=&\sigma^2_1/(\sigma_0\sigma_2)
\EA 
\BA\label{eq:GRFeq4}
\theta_*&=&\sqrt{2}\sigma_1/\sigma_2
\EA 
\BA\label{eq:GRFeq5}
\sigma^2_p&=&\nonumber\int^\infty_0\frac{\ell d\ell}{2\pi}\ell^{2p}P_\ell\\&=&p!2^{2p}(-1)^p\frac{d^p\xi}{d(\theta^2)^p}(0),
\EA 
and where $P_\ell$ is the continuous 2d power spectrum of the
convergence field, and $\xi(\theta)$ is its two-point correlation
function.  Integrating Eq.~(\ref{eq:GRFeq1}) over $\nu$ gives the
total number of peaks $n_{\rm pk}$ regardless of their height,
\BA\label{eq:GRFeq6} 
n_{\rm pk}&=&(4\pi\sqrt{3})^{-1}{\theta_*}^{-2}.
\EA

To verify the accuracy of these analytic formulae, we produced 200
numerical maps of GRFs, by generating 200 independent random
realizations of the theoretical 2d weak lensing power spectrum (in our
fiducial model, assuming a source redshift $z_s=2$).  We first
generate a 2d complex random field in $\ell$ space, with the real and
imaginary parts of Fourier modes distributed independently, following
Gaussians with a standard deviation of $\sqrt{P_\ell/2}$. Here
$P_\ell$ is the power spectrum \cite{BBKS}.  We then perform a
discrete Fourier transform to produce maps in real space.  The GRF
maps have a size of $2048\times2048$, to mimic the actual WL maps.
Noise is then added according to Eqs.~(\ref{eq:ellipnoise1})
and~(\ref{eq:ellipnoise2}), with $n_{\rm gal}$=15, and a 1 arcmin
smoothing is applied. To avoid edge effects, we discard pixels located
within 40 pixels ($\approx3\times$ the smoothing scale) of the map
edges.

To reproduce the random GRF realizations as closely as possible, we do
not calculate $\sigma_0$, $\sigma_1$, $\sigma_2$ from the input power
spectrum; instead, we measure these directly from the maps, using the
finite difference derivatives (ignoring the constant coefficients in
Eq.~(\ref{eq:GRFeq5}))
\BA\label{eq:measuresigma1}
\sigma^2_0&=&\langle \kappa^2\rangle-{\langle \kappa \rangle}^2
\EA 
\BA\label{eq:measuresigma2}
\sigma^2_1&=& \langle\left(\frac{d\kappa}{dx}-\overline{\frac{d\kappa}{dx}}\right)^2+\left(\frac{d\kappa}{dy}-\overline{\frac{d\kappa}{dy}}\right)^2 \rangle
\EA 
\BA\label{eq:measuresigma3}
\sigma^2_2&=& \langle\left(\frac{d^2\kappa}{dx^2}-\overline{\frac{d^2\kappa}{dx^2}}+\frac{d^2\kappa}{dy^2}-\overline{\frac{d^2\kappa}{dy^2}}\right)^2 \rangle
\EA 
where $\kappa$ is the 2d convergence field, and
$\overline{\frac{d\kappa}{dx}}, \overline{\frac{d\kappa}{dy}},
\overline{\frac{d^2\kappa}{dx^2}}$, and
$\overline{\frac{d^2\kappa}{dy^2}}$ are the averages of the
corresponding first and second derivatives.  For reference, this leads
to a prediction of 1680 peaks, compared to the actual number 1685
found in the noiseless maps; the prediction based on the analytic
calculation of the $\sigma$'s is slightly worse, 1649.  We have
checked that the situation is similar in the true WL maps: measuring
$\sigma_{0,1,2}$ numerically gives a slightly more accurate prediction
for the total number of peaks than calculating $\sigma_{0,1,2}$
analytically from the power spectrum through Eq.~(\ref{eq:GRFeq5}).

In Fig.~\ref{fig:GRFpeaks}, we show predictions from the analytic
formulae~(Eqs.~(\ref{eq:GRFeq1})-(\ref{eq:GRFeq5})), and the mean peak
counts in our 200 mock GRF maps.  This tests our reproduction of the
formulae, as well as the accuracy of our numerical measurements of
$\sigma_0$, $\sigma_1$, $\sigma_2$.  The agreement is excellent, with
residuals of only $\lsim~2$ percent over most of the range shown.

The main advantage of creating mock numerical realizations of the GRF
maps is that we can use these to measure the (co)variance in the
Gaussian peak counts. This covariance matrix is necessary to compute
the $\Delta\chi^2$ values between pairs of cosmologies in the Gaussian
case (see discussion below).

\subsection{Statistical Comparisons}
\label{subsec:statisics}

The basic statistical task in this paper is to assign a significance
of the difference between a pair of maps, given the stochastic
fluctuations in the maps over many realizations.  This is required in
order to quantify how well two cosmological models can be
distinguished with the peak counts.

The simplest statistical test consists of computing
$\Delta\chi^2_{f^\prime,f}$ between a pair of cosmologies $f$ and
$f^\prime$, using the {\em mean} number of peaks in each bin, averaged
over all realizations,
\BA\label{eq:meanchi2}
\Delta\chi^2_{f^\prime,f}&=&\nonumber\mathbf{dN}^{(f^\prime,f)}(C^{(f)})^{-1}\mathbf{dN}^{(f^\prime,f)}\\&=&\sum_{ij}dN_i^{(f^\prime,f)}(C^{(f)})_{ij}^{-1}dN_j^{(f^\prime,f)},
\EA 
where $dN_i^{(f^\prime,f)}\equiv \overline N_i^{(f^\prime)}-\overline
N_i^{(f)}$ is the difference between the average number of peaks in
bin $i$ in cosmology $f^\prime$ and in cosmology $f$.  Note that $i$
here can label bins of different peak--heights, but can also include
different source galaxy redshifts or smoothing scales.  Here $C^{(f)}$
denotes the covariance matrix of the number of peaks in cosmology $f$,
\BA\label{eq:meanchi3}
C^{(f)}_{ij}=\frac{1}{R-1}\sum_{r=1}^R(N^{(f;r)}_i-\overline{N}^{(f)}_i)(N^{(f;r)}_j-\overline{N}^{(f)}_j)
\EA 
where $N^{(f;r)}_i$ is the number of peaks in bin $i$ in the $r^{\rm
th}$ realization (i.e.\ convergence map) of the cosmology $f$, and $R$
is the total number of realizations.

This $\Delta\chi^2_{f^\prime,f}$ could be interpreted directly as a
likelihood or confidence level only if (i) the peak count distribution
in each bin were Gaussian, and (ii) the mean peak counts depended
linearly on the cosmological parameters.  As long as the change in
parameters is small, a Taylor expansion to the first order is a good
approximation, and therefore the second condition is unlikely to be
strongly violated in our case.  We will verify below that condition
(i) is satisfied to a good accuracy, as well. To be specific, we
examine directly the distribution of the quantity defined as
\BA\label{eq:chisq1}
\chi^2_{f^\prime}(r)&=&\sum_{ij}dN_i^{(f^\prime;r)}(C^{(f)})_{ij}^{-1}dN_j^{(f^\prime;r)}
\EA 
where $dN_i^{(f^\prime;r)}\equiv N_i^{(f^\prime;r)}-\overline
N_i^{(f)}$ is the difference between the number of peaks in bin $i$ in
realization $r$ of a test cosmology $f^\prime$, and the average number
of peaks in the same bin in the fiducial cosmology $f$.  We will show
that $\chi^2_{f}(r)$ closely follows a true chi-squared distribution.

Unless stated otherwise, in this paper, we use five $\kappa$ bins to
calculate $\Delta\chi^2$. The bin boundaries are chosen by visual
inspection, using two rough criteria: (i) the difference in the peak
height distributions in the two cosmologies should not change sign
within any of the bins, and (ii) the numbers of peaks in each bin
should be as comparable as possible.  The influence of the number of
bins and the bin boundaries on our results will be discussed in
\S~\ref{sec:discussion} below.

\subsection{Fisher Matrix and Marginalized Error }
\label{subsec:Fisher matrix}

With the assumptions that (i) the observables, i.e. the mean number of
peaks in each bin $\overline{N}_i$, depend linearly on the
cosmological parameters $p$; and (ii) that for a fixed $p$, the
probability distribution of $N_i$ follows a Gaussian, the marginalized
error on each parameter can be calculated using the Fisher matrix
(e.g. ref.~\cite{Tegmark+97}).

The Fisher matrix for point $p_0$ in the parameter space is given by
the matrix trace
\BA\label{eq:Fisher1}
F_{\alpha\beta}&=&\frac{1}{2}Tr[C^{-1}C_{,\alpha}C^{-1}C_{,\beta}+C^{-1}M_{\alpha\beta}],
\EA 
with
\BA\label{eq:Fisher2}
M_{\alpha\beta}&\equiv&\overline{N}_{,\alpha}\overline{N}^T_{,\beta}+\overline{N}_{,\beta}\overline{N}^T_{,\alpha},
\EA
where the Greek indices refer to model parameters, a comma preceding
an index denotes a partial derivative with respect to the
corresponding parameter, $C_{ij}$ is the covariance matrix of $N_i$'s,
and $\overline{N}_i$ is the expectation value of $N_i$.  In this
paper, we will consider only the second term in Eq.~(\ref{eq:Fisher1})
above.  The constraints through this term arise from the dependence of
the mean number of peaks $\overline{N}_i$ on the cosmological
parameters. In principle, additional constraints could be available
from the first term, which represents the dependence of the
(co)variances $C_{ij}=\langle N_i N_j\rangle$ on cosmology (see
refs.~\cite{LH04} and \cite{HK03} for related points in the context of
cluster counts). Our results suggest that this dependence is
relatively weak; in practice, however, we do not have a sufficient
number of independent realizations to accurately evaluate this
dependence.  From the Fisher matrix, the marginalized error on the
parameter $\alpha$ is calculated simply as $\sigma_\alpha =
(F^{-1})_{\alpha\alpha}^{1/2}$. (In contrast, the square root of
Eq.~(\ref{eq:meanchi2}) divided by the difference between model
parameters corresponds to the parameter sensitivity with all other
parameters fixed.)

\section{Results}
\label{sec:results}

\subsection{The Physical Origin of Peaks and of their Cosmology Dependence}
\label{subsec:halos}

In general, peaks in the convergence field can arise for different
reasons.  They could be caused by (i) one or more collapsed halos
along the LOS; (ii) large-scale, mildly overdense filaments, seen in
projection; (iii) non--linear ``protoclusters'' that are on their way
to collapse, but have not yet virialized and settled into an
equilibrium structure~\cite{WK02,DFN09}; and (iv) pure galaxy shape
noise.  In reality, peaks can be produced by a combination of the
above effects.

Relatively high--amplitude ($\gsim 3.5\sigma$) lensing peaks have been
studied thoroughly in the past (e.g. \cite{White+02,SMCWLS,HS05}), and
it is known that a large fraction of these peaks is attributable to a
single collapsed massive halo.  Our motivation to revisit this topic
is that the cosmological information is contained primarily in the
lower--amplitude peaks, whose origin has not yet been clarified.  As
described in section II, we start with each individual peak, and
identify all halos along the sightline.  Likewise, for each halo, we
identify peaks that are located within a 3 arcmin distance from the 2d
sky position of the halo.

We would also like to know why the number of peaks changes with
cosmology.  To help clarify this, we examine realizations of pairs of
models with different values of $\sigma_8$ with {\em quasi identical}
initial conditions.  In these pairs of models, we use the same random
seeds to generate the amplitudes and phases for the Fourier modes of
the density and velocity field at redshift $z=0$, and then scale these
back to the starting redshift of the simulation, with the linear
growth factor.  Since the power spectra differ only by an overall
normalization, and since the growth factor is independent of
$\sigma_8$, the initial conditions, as well as the final WL maps,
maintain very similar patterns.  Therefore we can attempt to match
individual peaks in the two cosmologies, to see what happens to a
given peak, when, say, $\sigma_8$ is increased.

Results in this section are based on 50 realizations of the noisy maps
in the fiducial and in the low-$\sigma_8$ models, with source galaxy
redshift $z_s=2$, $n_{\rm gal}=15$ and 1 arcmin smoothing.  The maps
have an angular size $3.46\times3.46~{\rm deg}^2$, $2048\times2048$
pixels.  To avoid edge effects, 30 pixels($\approx$ 3 times the
smoothing scale) from all four edges of each map are discarded,
leaving an area of $3.36 \times 3.36~{\rm deg}^2$.  The fiducial model
is used for studying the origins of the peaks, and a comparison
between the fiducial and the low--$\sigma_8$ model is used for
studying the cosmology-sensitivity.

We first perform an analysis of peak-halo matching, closely following
Hamana et al.~\cite{SMCWLS}, for both high and medium peaks.  High
peaks are defined as those with $\nu\ge4.8$, where $\nu$ is the
maximum value $\kappa_{\rm peak}$ of the peak, in units of the
standard deviation of the noise field.  We restrict our halo catalog
to halos that are expected (based on the NFW approximation;
Eqs.~(\ref{eq:NFW1})-(\ref{eq:convergence}) with $\phi_0=0$) to
produce a peak with a height of $\nu_{\rm nfw}\ge4.8$.  Similarly, for
medium peaks with $1.1 \le \nu < 1.6$, we restrict our halo catalog to
those halos with $1.1 \le \nu_{\rm nfw} < 1.6$.  We carry out the
peak-halo matching by searching for a matched pair candidate within a
radius of 1.8 arcmin from the peak position or from the halo center.
This maximum angular separation is chosen, as in~\cite{SMCWLS}, so
that it is larger than the smoothing radius of 1 arcmin, while it is
still smaller than the angular virial radius of a massive halo at
$z\lsim1.3$, where the lensing kernel has the most weight. If there is
more than one candidate pair within this radius, we
follow~\cite{SMCWLS} and adopt the closest one as the primary
candidate.

\begin{table}
\begin{tabular}{|c|c|cc|} 
\hline
class & matching result & \multicolumn{2}{c}{number of matches} \vline \\
 & & (high peaks) & (medium peaks)\\
 \hline
 $i$   & halo $\Leftrightarrow$ peak & 526 (0.93)  & 2653 (4.7)  \\
 $ii$  & halo with no paired peak    & 230 (0.41)  & 19609 (35) \\
 $iii$ & peak with no paired halo    & 2264 (4.0) & 24709 (44) \\
 $iv$  & halo $\Rightarrow$ peak     & 2 (0.0035)    & 90 (0.16)    \\
 $v$   & halo $\Leftarrow$ peak      & 12 (0.021)   & 194 (0.34)   \\
 \hline
\end{tabular}
\caption[]{\textit{Matching of halos and peaks similar to Hamana et
    al.~\cite{SMCWLS}, but shown separately for high and medium
    peaks. In the second column, ``$\Leftrightarrow$'' indicates a
    primary match in both directions, whereas $\Rightarrow$ and
    $\Leftarrow$ indicate a primary match in the direction of the
    arrow only.  In total, our 50 realizations of $3.36\times3.36
    deg^2$ maps contain 2,802 high peaks and 27,556 medium peaks,
    whereas the three--dimensional N-body outputs contain 758 massive
    halos and 22,352 medium halos with corresponding masses. The
    numbers in () show the number per $deg^2$ averaged over 50
    realizations, to be compared with the results of Hamana et
    al.~\cite{SMCWLS} (see text for discussion).}}
\label{tab:Hamana}
\end{table}

The 50 noisy maps contain a total of 2,802 high peaks and 27,556
medium peaks.  For comparison, the halo catalogs contain 758 massive
halos with $\nu_{\rm nfw} \ge 4.8$ and 22,352 medium-sized halos with
$1.1 \le \nu_{\rm nfw} < 1.6$.  Following Hamana et al.~\cite{SMCWLS},
we sort the results of the matches into the following five categories:
($i$) both a halo and a peak are each other's primary pair candidate;
($ii$) a halo without a paired peak ($iii$) a peak without a paired
halo; ($iv$) a halo has a matched peak, but is {\em not} the primary
matched halo of that peak; and ($v$) a peak has a matched halo, but it
is {\em not} the primary matched peak of that halo.
Table~\ref{tab:Hamana} shows the number of matches, separately for
high and medium peaks, falling into each category.

Bearing in mind differences in redshift, noise, and peak height
thresholds, our results are in reasonable agreement
with~\cite{SMCWLS}.  Overall, we have found, on average, 5.0
deg$^{-2}$ high peaks and 1.3 deg$^{-2}$ massive halos.  The majority
($70\%$) of the massive halos produce a one-to-one matching peak,
although these account only for $\approx 20\%$ of the total population
of high peaks.  More specifically, \cite{SMCWLS} finds 23 deg$^{-2}$
high peaks and 8.1 deg$^{-2}$ massive halos.  Their matches per
$deg^2$ in categories $i$, $ii$, and $iii$ are 5.9, 2.1, 15
accordingly.  In general, we have found fewer peaks and halos
than~\cite{SMCWLS}, which can be attributed to our lower $\sigma_8$
and $\Omega_m$ values. Our threshold of high peaks is also not
identical to theirs.  However, we have checked that the number of
halos above a threshold is in good agreement with their equation
15~\cite{SMCWLS}, when our cosmological parameters are used.  We have
found a slightly lower completeness of the halos (less fraction of
halos in category $i$ and more fraction in category $ii$), and also a
$\sim50\%$ lower purity (more fraction of peaks in category $iii$).
We have chosen a higher source redshift, $z_s=2$ (\cite{SMCWLS} used
$z_s=1$), this makes projection effects more important, and may
explain why our completeness is lower.  Our noise is slightly larger,
and our simulations and final maps have higher resolution, compared
to~\cite{SMCWLS} -- these effects tend to increase the number of peaks
relative to number of halos, and to reduce the purity of identifying
halos.

The most important result in Table~\ref{tab:Hamana} is that, in
difference from massive halos, only a small fraction ($12\%$) of the
22,352 medium-sized halos produce a medium peak with a one-to-one
pair, accounting for less than $10\%$ of all medium peaks.  We
conclude that the close agreement in the number of medium peaks and
halos (27,556 vs.  22,352) is a coincidence - and noise and projection
effects are much more important for medium peaks than for high peaks.

We next extend the above analysis, by identifying all halos, down to
low masses, by using a larger, 3 arcmin cone around the LOS toward
each peak, and by computing the expected contribution of each halo
$\kappa_i$ to the total convergence at the position of the peak (based
on the NFW approximation).  We then rank the halos according to their
$\kappa_i$ values (starting from highest and going down to the
lowest). We add the noise $\kappa_{\rm noise}$ at the peak location to
this ranked list, and ask the following question: starting from the
highest value, how far down this ranked list do we need to sum the
contributions, before they account for $>50\%$ of the total peak
height?

\begin{figure*}[htp]
\centering
\includegraphics[width=8 cm]{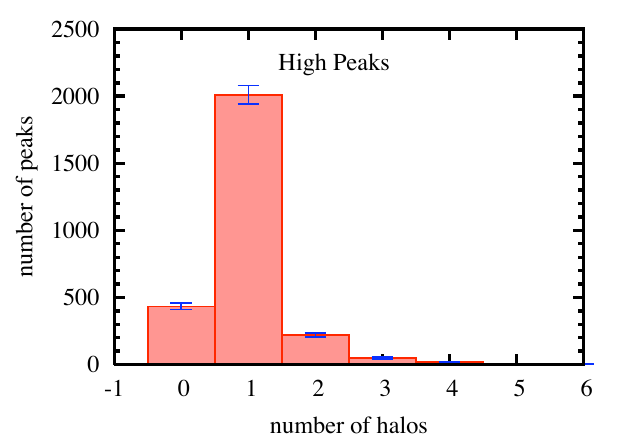} \includegraphics[width=8 cm]{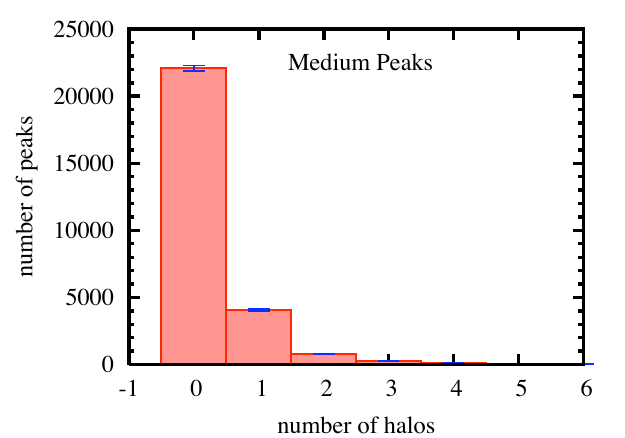}
\includegraphics[width=8 cm]{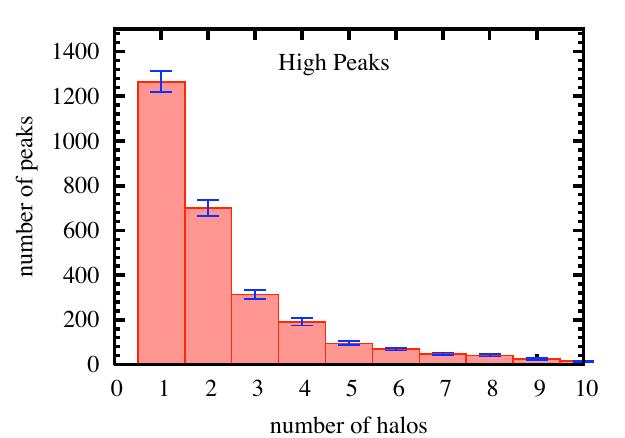} \includegraphics[width=8 cm]{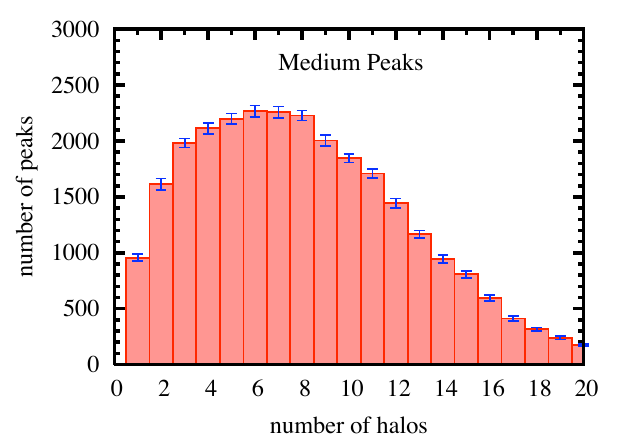}
\hfill
\caption[]{\textit{These figures illustrate the number of halos, as
    well as noise, contributing to medium and high peaks in 50
    realizations. For each peak, we identify all halos along the
    sightline, and rank them (as well as noise) according to their
    contribution to the peak convergence.  In the upper two panels, we
    show the distribution in the number of halos required to account
    for $>50\%$ of the total peak height (with ``0'' corresponding to
    cases in which noise alone explains half of the peak value).  In
    the lower two panels, noise is excluded, and we show the
    distribution in the number of halos required to account for
    $>50\%$ of the total halo contribution. Error bars are estimated
    as the the standard deviation of the number of counts in each bin,
    multiplied by $\sqrt{50}$.  }}
\label{fig:halocontributions}
\end{figure*}

In the upper two panels of Fig.~\ref{fig:halocontributions}, we plot
the distribution of this quantity; a ``0'' indicates that noise is the
single largest contributor, and already accounts by itself for most of
the peak, ``1'' indicates that at least 1 halo had to be included,
etc.  These panels clearly show that the large majority of high peaks
are dominated by a single halo, which accounts for at least half of
the peak amplitude. Most of these halos fall below the expected
threshold $\nu_{\rm nfw}=4.8$. The high peaks thus have a much better
one-to-one match with halos than Table~\ref{tab:Hamana} implies, once
lower--mass halos are included.  On the other hand, the large majority
of medium peaks are dominated by noise.

\begin{figure}[htp]
\centering
\includegraphics[width=8 cm]{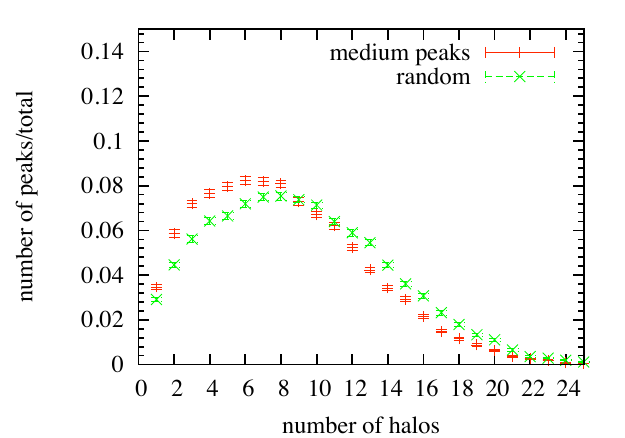} \hfill
\caption[]{\textit{As in the lower panels in
Figure~\ref{fig:halocontributions}, we show the distribution in the
number of halos required to account for $>50\%$ of the total halo
contribution. Here we contrast these distributions for medium-height
peaks and for randomly chosen directions on the sky.  The figure
demonstrates that medium--height peaks preferentially pick out
directions toward conjunctions of $\sim4-8$ halos. }}
\label{fig:randomdirections}
\end{figure}

Since noise does not contain any cosmological information, in the
bottom two panels of Fig.~\ref{fig:halocontributions}, we repeat the
same exercise, except that the noise contribution is excluded, and we
show the number of halos required to account for the total {\em halo}
contribution.  These panels show that while the high peaks are
typically dominated by a single halo, the contributions from a second
(or higher-rank) halos is often ($\gsim50\%$ of cases) as important.
In the case of medium peaks, however, it is very rare ($<5\%$ of
cases) for a single halo to dominate the cumulative halo
contribution. Instead, there is a broad distribution, but typically
(in $\sim$half the cases) $4-8$ halos are required to account for
$>$half of this total halo contribution.  As a sanity check, we have
computed the analogous distribution for random directions on our maps
(i.e., not toward peaks). For these random directions, as shown in
Figure~\ref{fig:randomdirections}, the distribution has an even
broader shape, centered at 8, and generally shifted toward larger
numbers of halos.  This reassures us that the medium peaks still do
preferentially pick out directions toward conjunctions of $\sim4-8$
halos.  We also examined the masses and redshifts of these dominant
$\sim4-8$ halos. We have found that the masses range between ${\rm
few}\times 10^{12}~{\rm M_\odot} < M < {\rm few}\times 10^{13}~{\rm
M_\odot}$, and found no correlations in redshift (i.e., the
contributing halos are not part of a single structure).  Finally, we
find that simply adding up the expected $\kappa$ contribution of {\em
all} halos along the LOS to a peak always overproduces the $\kappa$
value of the peak (not surprising, since this neglects the $\kappa$
deficit from underdense regions).

We next wish to clarify why there are fewer medium-height peaks when
$\sigma_8$ is increased (and vice versa). One can intuitively guess
that increasing $\sigma_8$ simply increases the "scatter" due to large
scale structures.  In the linear regime, changing $\sigma_8$ simply
changes the local (3d) density contrast, by the same factor
everywhere. Pretending that this holds in the nonlinear regime, it is
easy to see that the set of peaks would be invariant under changing
$\sigma_8$ -- however, positive peaks would be enhanced, and negative
peaks (i.e., maxima residing inside large-scale voids) would become
yet more negative.  This would broaden the peak-height distribution,
and reduce the number of peaks near $\kappa_{\rm peak}\sim 0$.  Of
course, this picture is over-simplified: as $\sigma_8$ is increased,
the change in the density field is not a simple re-scaling;
furthermore, peaks can be destroyed and new peaks can be created.
Indeed, we already know that the total number of peaks changes with
$\sigma_8$ (Paper I).

To understand the dominant effect, we attempt to follow and match
individual peaks in a pair of models with different $\sigma_8$.  We
have found that a direct matching of peaks is not possible, because
the locations of the peaks tend to shift (by several to more than 20
pixels) and therefore the correspondence between peaks in two
different maps remains ambiguous (except for the very highest and most
conspicuous peaks).  Instead, we proceed by using halos as
intermediate proxies for the peaks.  Starting from a peak in the first
cosmology, we identify the halo that contributes most to this peak.
For this analysis, we consider only those peaks for which the matched
primary halo contributes at least 10\% of the total halo contribution
(otherwise it is unfair to use the halo as a proxy for the peak).  We
next search the entire halo catalog in the second cosmology, and
identify the ``same" halo, by finding the one that shares most common
particles with the halo in the first cosmology.  Finally, we search
through the peaks in the second cosmology within a cone of 3 arcmin
around the halo, requiring that their height is within the range
$\pm\sigma_{\rm noise}$ of the original peak in the first cosmology,
and that the ``same" halo contributes at least 10\% of the total halo
contribution.  If no such peak is found, then the match is declared
unsuccessful.  If more than one such peak is found, we select the one
to which the matched halo contributes the most.

\begin{figure*}[htp]
\centering
\includegraphics[width=8 cm]{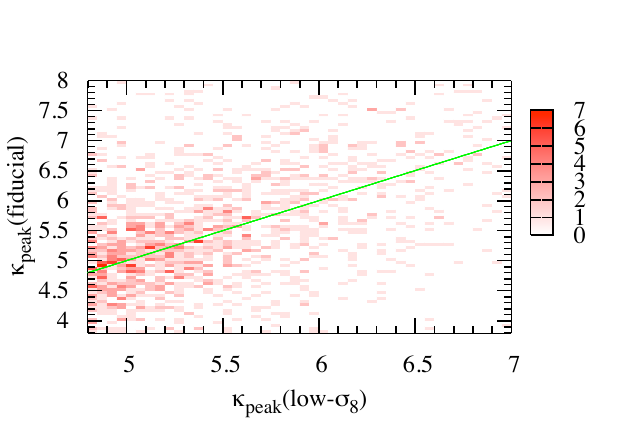}
\includegraphics[width=8 cm]{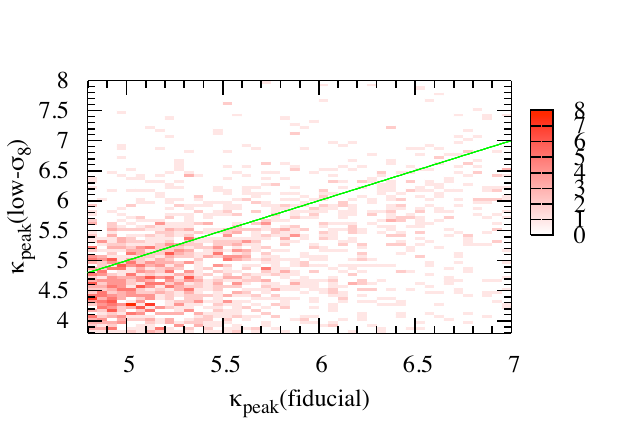}
\includegraphics[width=8 cm]{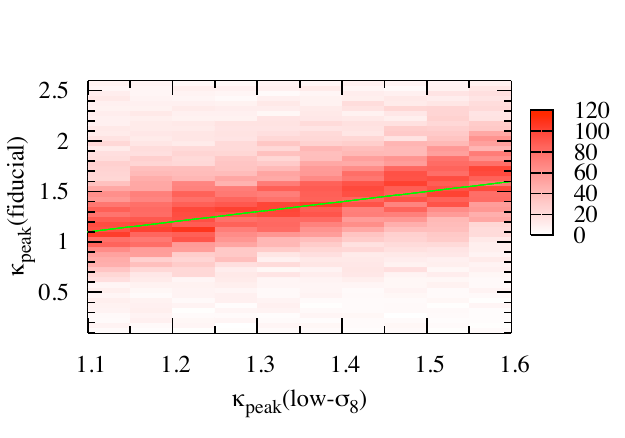}
\includegraphics[width=8 cm]{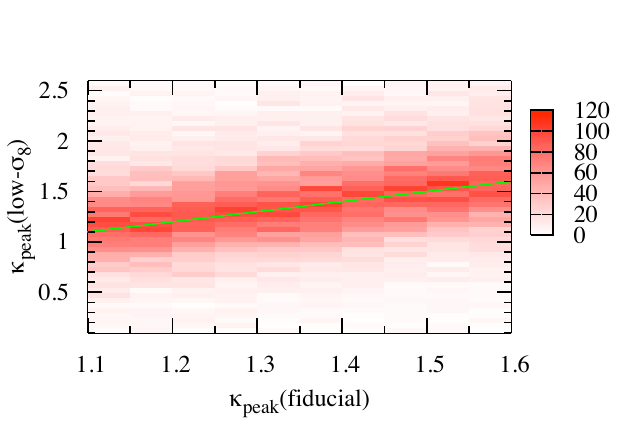}
\hfill
\caption[]{\textit{The figure shows the results of an attempt to match
    individual high and medium peaks in a pair of cosmologies, as
    explained in the text.  The x-axis shows the peak height in the
    starting cosmology model, and the y-axis shows the height of the
    ``same'' peak in the model to be matched with the starting model
    (whenever the peak has a match).  The top-left panel shows matches
    found in the fiducial model for the high peaks in the
    low-$\sigma_8$ cosmology (and vice versa in the top-right panel);
    the bottom two panels repeat the exercise for medium peaks.
    Typically $\sim 80\%$ of the high peaks have a match, but only
    $\sim50\%$ of the medium peaks do.}}\label{fig:peakmatching}
\end{figure*}

Fig.~\ref{fig:peakmatching} shows the results of the above matching
procedure between the fiducial and the low-$\sigma_8$ models, for both
high peaks (upper two panels) and medium peaks (lower two panels).
Focusing on the high peaks first, there are 1,987 high peaks in the
lower $\sigma_8$ model, and 2,802 high peaks in the fiducial model.
In the upper-left panel, we start with the peaks in the low-$\sigma_8$
model, and show their matches in the fiducial model. In the
upper-right panel, we reverse the direction, and start with the
fiducial model.  We find an ``unambiguous'' match (in the sense
defined in the preceding paragraph) for 87.1\% and 80.8\% of the
peaks, respectively.  Proceeding to the two lower panels, we show the
results for the medium peaks.  There are 29,097 medium-height peaks in
the low-$\sigma_8$ model; only 56.2\% of these have a matching peak in
the fiducial model.  Likewise, starting from the 27,556 medium-height
peaks in the fiducial model (bottom panel), we find that 55.7\% of
these have a match in the low-$\sigma_8$ cosmology.

In conclusion, most high peaks are matched to a peak in the other
cosmology, although the peak $\kappa$ values in the two cosmologies
differ by an amount comparable to $\sigma_{\rm noise}$.  In contrast,
about half of the medium-height peaks do not have a clear match in the
other cosmology.  This is despite the fact that we use the same
realization of the noise map in both cosmologies (i.e., we avoid
creating an entirely different set of peaks by a different
noise-realization), and despite our rather lenient definition of a
``match''.  We speculatively interpret this result as follows. Since
the medium peaks are typically created by the sum of pure noise and
many halos in projection, their existence and their amplitude are both
sensitive to small changes in the spatial distribution and masses of
these halos.  In indirect support of the above conclusion, we have
identified the following trend: on average, the proxy-halo contributes
49.6\% for high peaks that have a match, {\it vs.} 33.8\% for those
that do not.  Similarly, for medium peaks, the proxy halo contributes
25.1\% {\it vs.} 18.9\% for match {\it vs.} unmatched peaks.  This
shows that whenever the dominant halo accounts for a smaller fraction
of the peak $\kappa$, it is more "fragile" and is less likely to have
a match in the other cosmology.  Finally, we have found that when
$\sigma_8$ is increased, then the matched halo in the
higher-$\sigma_8$ cosmology has typically grown more massive, with an
increase by 15\% and 12\% on average for matched and unmatched high
peaks, 16\% and 14\% for matched and unmatched medium peaks.  This
trend, however, does not hold in the $\kappa$ values of the matched
peaks; as Fig.~\ref{fig:peakmatching} shows, the peak heights have a
significant scatter, but a relatively low bias, between the
cosmologies.  This could be explained by the fact that underdense
voids become even more underdense when $\sigma_8$ is increased, which
tends to ``cancel'' the increase in $\kappa_{\rm peak}$ caused by the
fattening of halos.  To be specific, we find a fractional bias
$\langle(\kappa_2-\kappa_1)/\kappa_1\rangle = 0.034$
and an r.m.s. scatter
$\langle(\kappa_2-\kappa_1)^2/\kappa_1^2\rangle^{1/2}= 0.19$.
when high peaks in lower $\sigma_8$ model are matched to high peaks in
fiducial model.  In the reverse direction, the bias and scatter are
-0.088 and 0.17. These results agree with the upper two panels in
Fig.~\ref{fig:peakmatching}, showing a positive bias when $\sigma_8$
is increasing, and a negative bias when $\sigma_8$ is decreasing.  The
bias for the medium peaks is less clear than for the high peaks.  In
fact, the bias is positive in both matching directions, with the
number for increasing $\sigma_8$ (0.068) more positive than the number
for decreasing $\sigma_8$ (0.032).  This is because the medium peaks
are dominated by noise, rather than halos, and the peaks which
increase in height because of the positive noise have a larger chance
to survive as a peak than those that are hurt by the noise.  The
scatter for medium peaks is $\approx0.32$, about twice the scatter for
high peaks.

\begin{table}
\begin{tabular}{|c|c|c|} 
\hline
class & low-$\sigma_8\rightarrow$ fiducial & fiducial $\rightarrow$ low-$\sigma_8$ \\
 \hline
 exit to low $\kappa$ & 3303 & 3420\\
 stay in bin & 7683 & 7515\\
 exit to high $\kappa$ & 5373 & 4408\\
 total matched & 16359 & 15343\\
 lost (unmatched) & 12738 & 12213\\
 \hline
\end{tabular}
\caption[]{\textit{We sort medium peaks into different categories,
based on the outcome of our attempt to find a match for each peak in
another cosmology.  The total number of medium peaks in the
low-$\sigma_8$ model and the fiducial model are 29,097 and 27,556
respectively.  While going from one cosmology to the other, peaks can
be ``lost'', they can stay within the same $\kappa$ bin, or they can
move out of the bin to higher or lower $\kappa$ values.  Additionally,
new medium peaks appear.}}
\label{tab: peakmatching_Npk}
\end{table}

Finally, we examine the ``movement'' of the peaks in height $\kappa$
as $\sigma_8$ is varied, in order to test our hypothesis, stated
above, that an increase in $\sigma_8$ tends to evacuate peaks from
near the $\kappa\sim 0$ (or near the maximum of the peak-height
distribution).  For example, we divide the 29,097 medium-height peaks
in the low-$\sigma_8$ model into several cases.  Approximately half
(12,738) of these peaks are unmatched: they ``disappear'' when
$\sigma_8$ is increased (equivalently, these are peaks that ``appear''
when one starts from the fiducial model, and decreases $\sigma_8$).
The remaining 16,359 matched peaks are further divided into middle,
lower and high cases, based on whether they remain in the original
medium-$\kappa$ bin (7,683), exit this bin toward higher $\kappa$
(5,373) or to lower $\kappa$ (3,303), in the other cosmology.  These
results, as well as the corresponding results in the reverse matching
direction, are summarized in Table~\ref{tab: peakmatching_Npk}.

By examining the table, we conclude there is indeed a preferentially
larger scatter in the direction out of the ``medium'' bin, when going
from the lower $\sigma_8$ to the fiducial model, compared to the
reverse direction: $(3,303+ 5,373) > (3,420+ 4,408)$.  This table
further reveals that there are two distinct reasons for the decrease
in medium-height peaks.  Approximately 2/3rd of the total decrease (of
29,097-27,556 = 1,541 $\approx$ 1500 peaks), or $\approx$ 900 peaks,
can be attributed to the above mentioned "scatter" due to the
increased density contrast -- i.e. more peaks moving out of the bin
than into the bin.  The remaining $\sim$1/3rd of the decrease is due
to losing peaks, i.e. $\approx$ 500 more peaks are destroyed than
created, as $\sigma_8$ is increased.  Based on the preceding
discussion, we speculate that this latter affect is caused by the
projections of multiple halos, which can create and destroy the
relatively low amplitude peaks.

In summary, the results in this section suggest that medium-height
peaks are almost always dominated by pure galaxy shape noise, but they
receive a significant contribution from collapsed halos, with
typically 4-8 halos in projection along the LOS. The halos drive the
cosmological sensitivity of these peaks in two ways: by (i) changing
the amplitudes of the noise peaks, and by (ii) destroying and creating
new peaks.  Between these last two effects, in the case of $\sigma_8$,
we found that the first is $\sim$twice as important as the second.

\subsection{Comparison to Gaussian Predictions}
\label{subsec:gaussianresults}

Our next task is to examine whether (i) the statistics of the peaks,
and (ii) their cosmology-sensitivity differs significantly from
predictions in a Gaussian random field. The degree of any departure
from a GRF is especially important to quantify for the medium peaks,
since the results of the last section suggest that these are heavily
dominated by pure Gaussian noise.

\begin{figure*}[htp]
\centering
\includegraphics[width=7 cm]{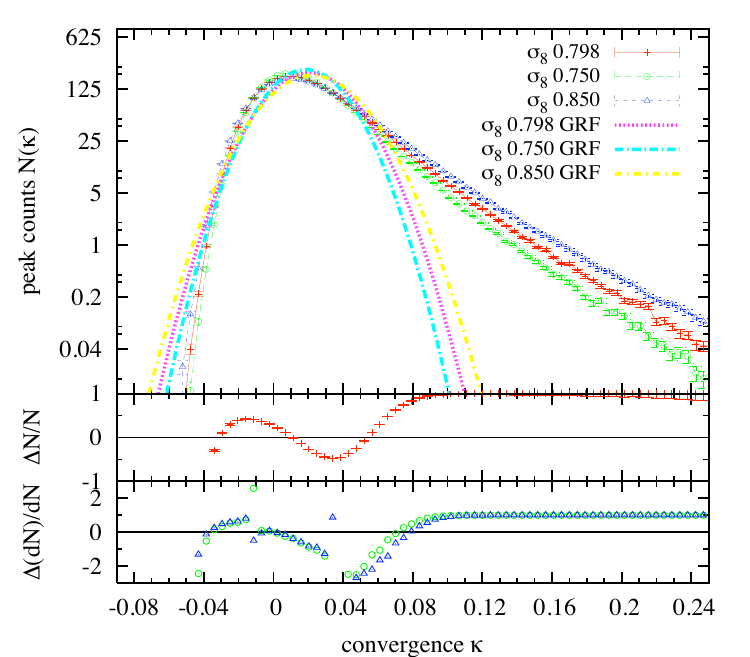}
\includegraphics[width=7 cm]{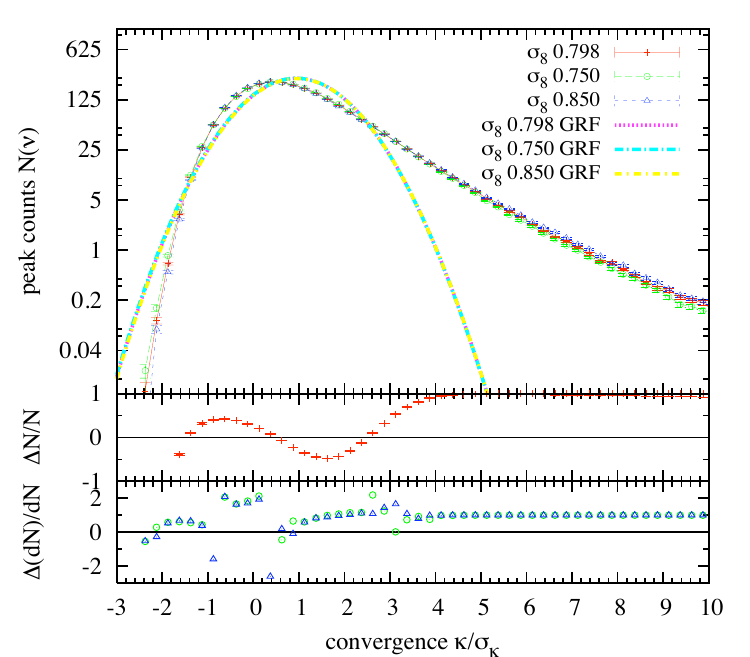} \\ \ \\
\includegraphics[width=7 cm]{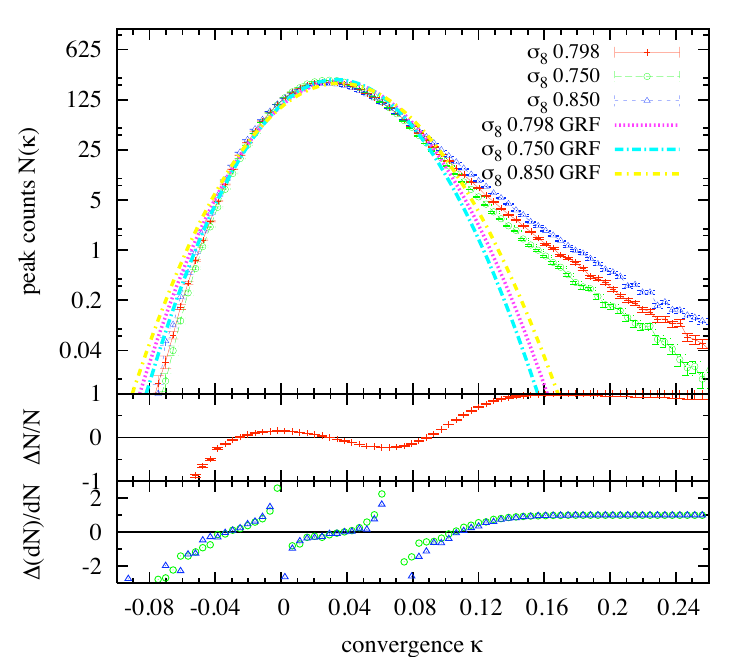}
\includegraphics[width=7 cm]{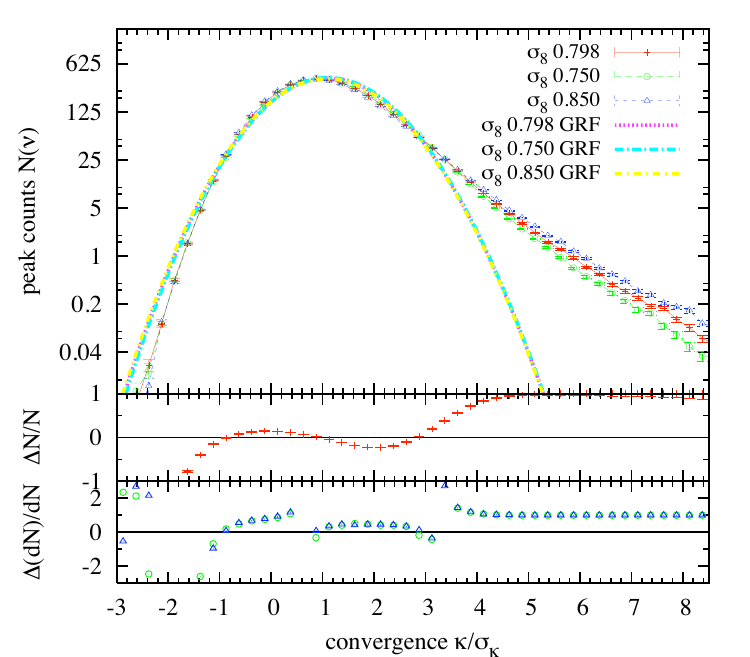}
\hfill
\caption[]{\textit{Number of peaks in our simulated $3.46 \times 3.46
    {\rm deg^2}$(including the 3 arcmin edge) convergence maps, in
    bins of width $\Delta\kappa$ = 0.0045.  In the right two panels,
    we have scaled the convergence field $\kappa$ by its r.m.s. value
    $\sigma_\kappa$; this removes information arising solely from
    $\sigma_\kappa$.  The source galaxies are assumed to be at $z_s =
    2$, and all results include 1 arcmin smoothing. In each panel, the
    three curves with data points correspond (from top to bottom on
    the right) to the high-$\sigma_8$, fiducial, and low-$\sigma_8$
    models.  The other three curves show theoretical predictions for
    peak counts in Gaussian random fields with the same three power
    spectra.
    In the bottom of each panel, we show the fractional difference
    between our fiducial model and a GRF for the peak counts
    $\langle(N_{simu}-N_{GRF})/N_{simu}\rangle$ (demonstrating that
    the WL peaks are strongly non-Gaussian), as well as for the
    difference in the peak counts between pairs of cosmologies
    $\langle(\Delta N_{simu}-\Delta N_{GRF})/\Delta N_{simu}\rangle$
    (explicitly demonstrating that the cosmology-sensitivity of our
    peak histograms is also different from that in a GRF).
    The four panels (upper-left, upper-right, lower-left, lower-right)
    show the results in noise--free unscaled, noise--free scaled,
    noisy unscaled and noisy scaled maps.}}\label{fig:peakcomparisons}
\end{figure*}

Our main results are shown in Fig.~\ref{fig:peakcomparisons}, which
directly compare the peak counts in our simulated maps with those in a
GRF.  The GRF predictions are computed from the theoretical formula as
discussed above, but using the (moments of) the power spectrum
$\sigma_{0}$, $\sigma_{1}$, $\sigma_{2}$ that were measured from the
corresponding simulated maps.  In each panel, we also compare the
high-$\sigma_8$, fiducial, and low-$\sigma_8$ models.  In the bottom
of each panel, we also show (i) the fractional difference between the
GRF and the fiducial model and (ii) how the change in the peak counts
between pairs of cosmologies is different in the GRF and our simulated
maps.  The source galaxies are assumed to be at $z_s = 2$, and all
results shown in the figure include 1 arcmin smoothing.  We plot the
mean number of peaks in convergence bins of width
$\Delta\kappa=\frac{1}{4}\sigma_{\rm noise}= 0.0045$, averaged over
1000 realizations.  In the top two panels, we exclude noise from the
maps; in the bottom two panels, noise is included.  Finally, in the
right two panels, we have scaled the convergence field $\kappa$ by its
r.m.s. value $\sigma_\kappa$ (these histograms use a bin width of
$\Delta (\kappa/\sigma_\kappa)= 0.25$).  This removes information that
arises from $\sigma_\kappa$ alone.  If the sole effect of changing
$\sigma_8$ was to change the heights of individual peaks by a constant
factor, then this would result in a re-scaling of the peak-height
probability distribution; the re-scaling by $\sigma_\kappa$ clarifies
the relative importance of this effect.

As these figures show, in the noiseless case, the peak height
distributions are very different from the Gaussian predictions and are
reminiscent of the skewed one-point function of $\kappa$, which has a
sharp drop at low demagnification, and a long tail to high
magnification~(e.g.~ref.~\cite{Wiley+09} and references therein).
This correspondence of the high-tails is not entirely surprising;
indeed, a pixel with a very high $\kappa$ value is likely to mark a
peak.  When noise is added, the distributions near their peaks look
much more similar to the Gaussian predictions.  However, there is
still a large non-Gaussian deficit of the lowest peaks (with the most
negative $\kappa_{\rm peak}$) and a clear excess of the highest
($\kappa \gsim 3\sigma_\kappa$) peaks.  Importantly, however, there
also remains a clear difference in the peak-height distributions even
for the medium-height ($\kappa \sim \sigma_\kappa$) peaks.  Finally,
as illustrated in the bottom insert in each panel, the
cosmology-sensitivity of our peak histograms is also different from
that in a GRF. These last points are encouraging, and suggest that the
medium peaks do contain non-Gaussian information.

\section{Discussion}
\label{sec:discussion}

\subsection{Sensitivity to Cosmological Parameters}
\label{subsec:Cosmology Sensitivity}

\begin{table}
\begin{tabular}{|c|c|cccc|} 
\hline
map & cosmology & \multicolumn{4}{c}{boundary} \vline\\
type & pair & \multicolumn{4}{c}{locations} \vline\\
 \hline
 noisy us & F and High-$\sigma_8$ & -0.0028 & 0.0217 & 0.0407 & 0.0695 \\
 noisy sc & F and High-$\sigma_8$ & 0.2650 & 0.6682 & 1.3550 & 3.3013 \\
  \hline
 noisy us & F and High-$\Omega_{m}$ & -0.0050 & 0.0200 & 0.0383 & 0.0627 \\
 noisy sc & F and High-$\Omega_{m}$ & 0.4618 & 0.9950 & 1.5750 & 3.0556 \\
  \hline
 noisy us&  F and High-$w$ & -0.0019 & 0.0190 & 0.0347 & 0.0551 \\
 noisy sc & F and High-$w$ & 0.2565 & 1.1450 & 2.4939 & 3.0368 \\
 \hline
\end{tabular}
\caption[]{\textit{Examples of bin boundaries used for the convergence
peak counts. The boundaries are listed for unscaled (``us'') and
scaled (``sc'') noisy maps, used to compute $\Delta\chi^2$ between the
fiducial model and the high-$\sigma_8$, high-$\Omega_{m}$, and
high-$w$ models, respectively.  In the unscaled case, the boundary
locations are in units of the dimensionless convergence $\kappa$; in
the scaled case, they are in units of $\nu=\kappa/\sigma_\kappa$.}}
\label{tab:chisq Simu Z2 boundaries}
\end{table}

\begin{table}
\begin{tabular}{|c|c|cccc|} 
\hline
map & cosmology & \multicolumn{2}{c}{noiseless $\Delta\chi^2$} & \multicolumn{2}{c}{noisy $\Delta\chi^2$}  \vline\\
type & pair & unscaled & scaled & unscaled & scaled \\
 \hline
 Sim & F and High-$\sigma_8$ & 5.16 & 0.46 & 5.89 & 4.29 \\
 GRF & F and High-$\sigma_8$ & 10.65 & 0.23 & 5.87 & 3.16 \\
  \hline
 Sim &F and Low-$\sigma_8$ & 5.01 & 0.34 & 5.09 & 3.67 \\
 GRF &F and Low-$\sigma_8$ & 9.93 & 0.16 & 4.98 & 2.58 \\
  \hline
 Sim & F and High-$\Omega_{m}$ & 3.61 & 0.033 & 4.02 & 2.46 \\
 GRF & F and High-$\Omega_{m}$ & 7.68 & 0.014 & 3.77 & 2.01 \\
  \hline
 Sim &F and Low-$\Omega_{m}$ & 4.39 & 0.053 & 4.44 & 2.56 \\
 GRF &F and Low-$\Omega_{m}$ & 8.79 & 0.043 & 4.08 & 2.11 \\
  \hline
 Sim &F and High-$w$ & 0.98 & 0.47 & 0.65 & 0.27 \\
 GRF & F and High-$w$ & 0.93 & 0.017 & 0.46 & 0.14 \\
  \hline
 Sim & F and Low-$w$ & 0.44 & 0.27 & 0.36 & 0.16 \\
 GRF &F and Low-$w$ & 0.54 & 0.004 & 0.26 & 0.08 \\
 \hline
\end{tabular}
\caption[]{\textit{$\Delta\chi^2$ values from our simulated maps and
from predictions in a GRF, based on the difference in the peak height
distributions between the fiducial model and six other models, varying
$\sigma_8$, $w$, and $\Omega_m$.  Results are shown for both the
unscaled ($N(\kappa)$) and the scaled ($N(\nu)$) peak distributions.
Source galaxies are assumed to be at $z_s=2$, and a set of
$2\times1000$ maps are used in comparing each pair of cosmologies.}}
\label{tab:chisq Simu GRF Z2}
\end{table}

The number counts have been found (Paper I) to depend sensitively on a
combination of ($\sigma_8$,$w$).  Here we vary $\sigma_8$ and $w$
separately, in order to clarify the sensitivity to each of these
parameters; we also consider variations in $\Omega_m$. We use
$\Delta\chi^2$, defined in Eq.~(\ref{eq:meanchi2}) above, to measure
the significance of the difference in the peak counts $N(\kappa)$,
caused by the changes in these parameters.  We used fiducial and other
cosmological maps to calculate the change in $N(\kappa)$, but we used
the control maps to compute the covariance matrix.  Having 9 times
more strictly independent realizations (45 control maps vs 5
realizations in the fiducial model) allows us to compute the
covariance matrix more accurately.  To isolate the sensitivity from
beyond a change in the r.m.s.  $\sigma_\kappa$, we also compute the
$\Delta\chi^2$'s between the scaled peak height distributions
$N(\nu)$.  In these analyses, we use five convergence bins whose
locations are chosen by visual inspection, as explained
above. Examples of bin boundaries we used are listed in
Table~\ref{tab:chisq Simu Z2 boundaries}.

Our main results are shown in Table~\ref{tab:chisq Simu GRF Z2}, and
can be enumerated as follows.

{\em Raw cosmology sensitivity.} The simulated noisy $\Delta\chi^2$
values in the unscaled maps are significant ($\Delta\chi^2\sim 4-6$),
and suggest that the cosmological sensitivity of the peak counts is
competitive with other methods (after scaling to the full size of an
all-sky survey, such as LSST; this extrapolation is discussed further
below).  The sensitivity for $w$ is about an order of magnitude weaker
($\Delta\chi^2\sim 0.3-0.6$) than for the other parameters. However,
this is the case for other observables, such as the power spectrum, as
well. As shown below (see Table~\ref{tab:merr} and related discussion)
the peak counts and the power spectrum individually have similar
sensitivities to all three parameters; they can furthermore be combined
to improve the marginalized errors by a factor of $\approx$two on all
three parameters.

{\em Can we ``scale out'' the cosmological information?} By comparing
the scaled and unscaled cases in the noisy maps, we see that scaling
the maps by the variance $\sigma_\kappa$ reduces the $\Delta\chi^2$
values only by a modest amount.  In these maps, only a small fraction
of the parameter-sensitivity arises through changes in
$\sigma_\kappa$.  Interestingly, the situation is different in the
raw, noiseless maps. Nearly {\em all} of the sensitivity in these maps
are attributable $\sigma_\kappa$: the $\Delta\chi^2$ values diminish
significantly after the scaling.  This result is somewhat
counter-intuitive, and implies that there is a ``non-linear''
interaction between noise and physical structures. More precisely, the
result can be re-stated as follows: before adding noise, the
cosmology-induced changes are very similar to a uniform 'stretching'
of the peak height distribution along the $x$-axis.  However, once the
noise is added, the cosmology-induced changes are no longer described
by such stretching.  In hindsight, this is not entirely surprising:
given that noise has almost no effect on the highest peaks, and has
increasingly larger impact on the lower peaks, it is to be expected
that the addition of noise spoils the ``linear'' stretching.

{\em The impact of noise.} We find, furthermore, that the addition of
the noise {\em increases} the $\Delta\chi^2$ values for $\sigma_8$ and
for $\Omega_m$, while for varying $w$, noise hurts.  This is similar
to a result we found in Paper I, namely that noise increases the
change in the number of peaks.  While in the unscaled maps, the
increase in $\Delta\chi^2$ is modest, in the scaled maps, the increase
is very significant (the interpretation of this is already explained
in the preceding paragraph).  This result -- i.e. that the addition of
pure noise helps increase the $\Delta\chi^2$ -- is also somewhat
counter-intuitive, and will be discussed in detail in
\S~\ref{subsec:noise} below.

{\em Which peaks drive the sensitivity?} To answer this question, we
calculate $\Delta\chi^2$ values separately for peaks with low, medium
and high amplitudes.  In the scaled maps, the low range is chosen to
be below -0.8$\sigma_\kappa$.  The medium range is
$\pm0.5\sigma_\kappa$ wide, centered on the mode of the peak height
distribution.  The high range is defined to be above the height at
which the peak-height distributions start to differ significantly in
any given pair of cosmology. This is inferred visually from
logarithmic scaled peak height distributions, such as those shown in
the right two panels in Fig.~\ref{fig:peakcomparisons},and typically
falls at $\sim 3.5\sigma_\kappa$, or $\kappa\sim 0.1$.  On the
unscaled maps, we use the same boundaries as above, converted to
$\kappa$ values using the fiducial model.  Both in the low and high
ranges, we adopt two (adjacent) bins, and in the medium range, we use
either two or three bins (depending on where the peak-height
distributions in a given pair of cosmologies
cross). Table~\ref{tab:chisq meidian and high boundaries} summarizes
the bin boundaries.

\begin{table}
\begin{tabular}{|c|c|cccc|} 
\hline
peak & cosmology & \multicolumn{4}{c}{$\kappa$-bin} \vline\\
type & pair & \multicolumn{4}{c}{boundaries} \vline\\
 \hline
 noisy low & F and High-$\sigma_8$ & -0.2697 & -0.031 & -0.0250 & \\
 noisy medium& F and High-$\sigma_8$ & 0.0150 & 0.0214 & 0.0390 & 0.0460 \\
 noisy high& F and High-$\sigma_8$ & 0.1070 & 0.1240 & 1.4000 & \\
  \hline
 noisy low & F and High-$\Omega_{m}$ & -0.2697 & -0.031 & -0.0250 &  \\
 noisy medium & F and High-$\Omega_{m}$ & 0.0156 & 0.0313 & 0.0469 & \\
 noisy high & F and High-$\Omega_{m}$ & 0.1000 & 0.1150 & 1.4000 & \\
 \hline
\end{tabular}
\caption[]{\textit{Bin boundaries used in the analysis to identify the
relative importance of low, medium, and high peaks. The boundaries
(including the end-points) are listed for the noisy $\Delta\chi^2$
between the fiducial model and the high-$\sigma_8$ and
high-$\Omega_{m}$ models, in units of $\kappa$.  Two adjacent bins are
used in the low and high ranges and either two or three bins in the
medium range.}}
\label{tab:chisq meidian and high boundaries}
\end{table}

\begin{table}
\begin{tabular}{|c|c|cccc|} 
\hline
peak & cosmology & \multicolumn{2}{c}{noiseless $\Delta\chi^2$} & \multicolumn{2}{c}{noisy $\Delta\chi^2$}  \vline\\
type &  & unscaled & scaled & unscaled & scaled \\
 \hline
 low & F and High-$\sigma_8$ & 0.62 & 0.035 & 0.32 & 0.025\\
 medium & F and High-$\sigma_8$ & 2.55 & 0.024 & 3.30 & 0.87\\
 high & F and High-$\sigma_8$ & 2.85 & 0.069 & 2.13 & 0.36 \\
 frac.  & F and High-$\sigma_8$ & 1.17 & 0.28 & 0.98 & 0.29 \\
  \hline
 low &F and Low-$\sigma_8$ & 0.34 & 0.05 & 0.21 & 0.02 \\
 medium &F and Low-$\sigma_8$ & 2.92 & 0.04 & 3.00 & 0.62 \\
 high & F and Low-$\sigma_8$ & 1.73 & 0.09 & 1.27 & 0.32 \\
 frac.  & F and Low-$\sigma_8$ & 1.00 & 0.52 & 0.88 & 0.26 \\
 \hline
 low & F and High-$\Omega_{m}$ & 0.53 & 0.009 & 0.23 & 0.01 \\
 medium & F and High-$\Omega_{m}$ & 1.53 & 0.004 & 2.65 & 0.70 \\
 high & F and High-$\Omega_{m}$ & 1.30 & 0.01 & 0.94 & 0.08 \\
 frac.  & F and High-$\Omega_{m}$ & 0.93 & 0.68 & 0.95 & 0.33 \\
\hline
low &F and Low-$\Omega_{m}$ & 0.36 & 0.01 & 0.25 & 0.025 \\
 medium &F and Low-$\Omega_{m}$ & 2.16 & 0.007 & 3.00 & 0.70 \\
 high & F and Low-$\Omega_{m}$ & 1.04 & 0.003 & 0.79 & 0.093 \\
 frac.  & F and Low-$\Omega_{m}$ & 0.81 & 0.36 & 0.91 & 0.32 \\
 \hline
\end{tabular}
\caption[]{\textit{$\Delta\chi^2$ values arising separately from peaks
in the low, medium and high range, with bin boundaries as specified in
Table~\ref{tab:chisq meidian and high boundaries}.  The fiducial model
is compared to models varying $\sigma_8$ and $\Omega_m$.  The 4$^{\rm
th}$ (last) row in each case shows the sum of the low, medium and high
$\Delta\chi^2$'s divided by the total $\Delta\chi^2$ obtained
previously and listed in Table~\ref{tab:chisq Simu GRF Z2}.}}
\label{tab:chisq meidian and high}
\end{table}

We show the resulting $\Delta\chi^2$ values from each type of peak in
Table~\ref{tab:chisq meidian and high}, for models varying $\sigma_8$
and $\Omega_m$.  As the difference caused by $w$ is small, we do not
discuss it here.  We also list the ratio $[\Delta\chi^2({\rm
low})+\Delta\chi^2({\rm medium})+\Delta\chi^2({\rm high})]/
\Delta\chi^2({\rm tot})$, where the numerator refers to the values
calculated here, and the denominator to the total $\Delta\chi^2$
computed above from the entire $\kappa$ range.  Even though the
low/medium/high ranges we use are disjoint, this ratio can exceed
unity (if in the original $\Delta\chi^2$, the bins were non-ideally
placed). The table shows that in the noisy maps, by far the largest
contribution comes from peaks in the medium range.  These are followed
in importance by the high and the low peaks.  We also see that in the
noisy unscaled case, the low, medium and high ranges together account
for essentially all ($>88\%$) of the total unscaled $\Delta\chi^2$.
In the scaled case, they add up to a smaller fraction
($\approx30-60\%$) of the total.

{\em How robust are the results?}  One may ask whether the
$\Delta\chi^2$ values (e.g. listed in Table~\ref{tab:chisq Simu GRF
Z2}) are robust under changes of the random realizations of the
underlying maps. This is a potential concern especially when the
$\Delta\chi^2$ values are low.  We used our control maps to re-compute
both the covariance matrix, and the change in $N(\kappa)$, and to see
how the $\Delta\chi^2$ values change.  We found values of
$\Delta\chi^2 > 1$ are very stable, and change by $< 5\%$. For $0.1 <
\Delta\chi^2 < 1.0$, the change is $\sim20\%$, and for the smallest
$\Delta\chi^2 < 0.05$, (occurring in scaled noiseless maps), the
change is $\sim50\%$.  Our finite number of realizations is therefore
only adequate to give an order-of-magnitude estimate of the
distinction between these pairs of maps.  We note that increasing the
number of bins also changes the $\Delta\chi^2$ values (increasing them
by $\sim 10\%$; see section~\ref{subsec:usebin}), but this change is
systematic, and does not influence our conclusions.

\subsection{Distinction from a Gaussian Random Field}
\label{subsec:DGRF}

We next turn to the question of whether the cosmology sensitivity
offers information beyond a pure GRF.  We know that the high peaks are
non-Gaussian, whereas the medium peaks, which drive the sensitivity,
appear to follow the GRF predictions more closely (though still
visibly deviate from them, even in the noisy maps).

We begin by directly quantifying the difference between the GRF and
the simulated peak-height distributions (shown in
Fig.~\ref{fig:peakcomparisons}).  In Table~\ref{tab:chisqGRFandSimu},
we show the $\Delta\chi^2$ values between the maps and the
corresponding Gaussian predictions.  In order to isolate the
non-Gaussian effects in the mean peak counts, the covariance matrix is
evaluated in the fiducial model (i.e. the covariance matrix in the GRF
is not used). The large numbers in this table reveal that overall, the
peaks are highly non-Gaussian, even in the noisy case.  Similar to
Table~\ref{tab:chisq meidian and high}, we study the significance of
the non-Gaussianity separately for low, medium and high $\kappa$
peaks.  We chose the ranges and the bin boundaries by the same
procedure as described for Table~\ref{tab:chisq meidian and high}.
Our results are summarized in Table~\ref{tab:chisqGRFandSimu regions},
and quantify the expectation that both the medium and high peaks
differ significantly from the GRF predictions, even in the noisy maps
(although the noiseless maps are more non-Gaussian).  In the noisy
maps, the significance of the non-Gaussianity for the low peaks is
relatively low, but this is likely a result of the relatively small
number of these low peaks.  The table also shows that the three
disjoint regions together account only for about 20-30\% of the total
$\Delta\chi^2$. This implies that the peak height distribution departs
from the GRF prediction everywhere (and no $\kappa$ values are
unimportant for the total $\Delta\chi^2$).

\begin{table}
\begin{tabular}{|c|c|c|} 
\hline
Model & noiseless $\Delta\chi^2$ & noisy $\Delta\chi^2$ \\
\hline
Fiducial & 164.20 & 44.98 \\
High-$\sigma_8$ & 191.22 &  64.47\\
Low-$\sigma_8$ & 130.20 &  32.23\\
High-$w$ & 177.85 &  47.91\\
Low-$w$ & 157.33 &  46.06\\
High-$\Omega_{m}$ & 180.53 &  55.14\\
Low-$\Omega_{m}$ & 146.27 &  36.15\\
\hline
\end{tabular}
\caption[]{\textit{$\Delta\chi^2$ values derived from the peak height
distributions between simulated maps and corresponding GRF
predictions. Source galaxies are assumed to be at $z_s = 2$, and 1000
noise free or noisy maps are used for each cosmological model.}}
\label{tab:chisqGRFandSimu}
\end{table}

\begin{table}
\begin{tabular}{|c|c|cc|} 
\hline
peak & cosmology & $\Delta\chi^2$ &  $\Delta\chi^2$ \\
type & pair & (noiseless)  & (noisy)  \\
 \hline
 low & Fiducial & 5.94 & 1.01\\
 medium & Fiducial & 9.56 & 3.08\\
 high & Fiducial & 17.94 & 9.79 \\
 frac.  & Fiducial & 0.20 & 0.31 \\
  \hline
 low & High-$\sigma_8$ & 7.43 & 1.68\\
 medium & High-$\sigma_8$ & 6.76 & 3.40\\
 high & High-$\sigma_8$ & 31.13 & 18.55 \\
 frac.  & High-$\sigma_8$ & 0.24 & 0.37 \\
  \hline
 low &Low-$\sigma_8$ & 3.25 & 0.59 \\
 medium &Low-$\sigma_8$ & 13.41 & 2.78 \\
 high &  Low-$\sigma_8$ & 10.10 & 5.20 \\
frac. & Low-$\sigma_8$ & 0.21 & 0.27 \\
 \hline
 low & High-$w$ & 5.78 & 1.05\\
 medium & High-$w$ & 9.98 & 2.72\\
 high & High-$w$ & 17.65 & 9.98 \\
frac. & High-$w$ & 0.19 & 0.29 \\
  \hline
 low & Low-$w$ & 6.05 & 1.12 \\
 medium & Low-$w$ & 6.91 & 2.48 \\
 high &  Low-$w$ & 18.34 & 9.73 \\
frac. &  Low-$w$ & 0.20 & 0.29 \\
 \hline
 low &  High-$\Omega_{m}$ & 6.87 & 1.39 \\
 medium & High-$\Omega_{m}$ & 6.78 & 3.45 \\
 high &  High-$\Omega_{m}$ & 26.45 & 15.23 \\
frac. & High-$\Omega_{m}$ & 0.22 & 0.36 \\
\hline
low & Low-$\Omega_{m}$ & 4.27 & 0.82 \\
 medium & Low-$\Omega_{m}$ & 14.03 & 3.21 \\
 high &  Low-$\Omega_{m}$ & 11.36 & 6.09 \\
frac. & Low-$\Omega_{m}$ & 0.20 & 0.28 \\
 \hline
\end{tabular}
\caption[]{\textit{$\Delta\chi^2$ values of peak height distribution
in low, medium and high range, between simulation and GRF theoretical
formula. Rate shows sum of low, medium and high $\Delta\chi^2$ divided
by the total $\Delta\chi^2$. Source is at z = 2.}}
\label{tab:chisqGRFandSimu regions}
\end{table}

The above demonstrates that one can (easily) tell the difference of
each map from a GRF.  We next ask whether the cosmology-induced
changes also differ from those in the Gaussian case.  To answer this,
we first calculate the $\Delta\chi^2$ between a pair of cosmologies,
using the expectation values of the GRF peak counts in both
cosmologies, computed from Eq.~(\ref{eq:GRFeq1}), with
$\sigma_{0,1,2}$ derived from the simulated maps in the corresponding
models.  These results are listed below the cosmological
$\Delta\chi^2$ values in Table~\ref{tab:chisq Simu GRF Z2}.  In order
to isolate non-Gaussian effects in the mean peak counts from those in
the covariance matrix, in these GRF calculations, we again use the
covariance matrix from the simulated fiducial model.  Therefore, these
GRF $\Delta\chi^2$ values do {\em not} represent the absolute
distinguishability of the two GRF maps; they are meant only to be
compared to the $\Delta\chi^2$'s from the corresponding cosmological
simulations.

As the comparisons of two adjacent rows in the table shows, the noisy
$\Delta\chi^2$'s in the simulations are generally close, overall, to
the corresponding values predicted in the GRF. This, of course, does
not necessarily mean that the information is the {\em same} as in a
GRF - indeed, we found above that the mean counts deviate from the GRF
predictions even in the noisy case.  However, the cosmology-induced
differences in the peak counts are, apparently, similar in magnitude
to that in a GRF.  When we repeat the GRF calculations with the
covariance matrix adopted from the mock GRF maps, we find that,
typically, the $\Delta\chi^2$ values increase by about a factor of
$\sim$4.  We find that in the GRF case, the standard deviation in the
peak counts in each bin is close to Poisson shot noise $\sim\sqrt{N}$.
In our maps, the fluctuations are typically larger, by up to a factor
of $\sim$4, which explains the corresponding reduction in the
$\Delta\chi^2$.  Interestingly, the variance in the 3D space density
of $\sim10^{14}~{\rm M_\odot}$ clusters has been found to exceed
Poisson noise by a similar factor~\cite{HK03}, providing physical
intuition for this result.

Finally, comparing the simulated and GRF--predicted values in the
noiseless case, we see that non-Gaussianity {\em reduces} the
significance between cosmologies before scaling, but {\em increases}
the significance after scaling.  This result makes sense: the peak
counts in the Gaussian case follow an almost strict linear scaling
with $\sigma_\kappa$, hence much of the difference disappears after
such scaling (although the linear scaling is not, in fact, exact, see
Eqs.~(\ref{eq:GRFeq1})-(\ref{eq:GRFeq2})).

\subsection{Information Beyond the Power Spectrum}
\label{subsec:Beyond power spec}

In the last section, we found that the peak height distribution is
very different from the expectation in a GRF with the same power
spectrum, especially for high peaks.  While this is encouraging, we
next study directly how much information is beyond the power spectrum.
This is important, since in general, a random field can be
non--Gaussian, but could still be fully characterized by its power
spectrum. For example, one can imagine that the 3--point (and higher
order) correlation functions are pre-specified functions of the power
spectrum.

We here measure the 2d convergence power spectrum
$\ell(\ell+1)P(\ell)$ directly from the map produced in each
realization, and treat it as another observable, in addition to the
peak height distribution.  The factor $\ell(\ell+1)$ is included to
make the observables in each bin close in magnitude (which helps make
matrix inversion more stable).  To be more specific, we first computed
the power spectrum in 200 equal--sized finer bins, with width
$\Delta\ell=531$.  The power was evaluated by taking the Fourier
transform of the convergence field and averaging the power in each
finer bin in the radial direction.  To calculate the $\Delta\chi^2$'s,
we considered only the range $100 < \ell < 20,000$, and divided this
range into 5 equal-sized bins (linear in $\ell$), and assigned each of
the 200 finer bins into one of these 5 bins (for computing
marginalized errors, we use 15, rather than 5 bins; see below).  Using
these five equally-spaced bins, we evaluate the mean
$\overline{P_i}\equiv\overline{\ell_i(\ell_i+1)P_i(\ell_i)}$ within
each bin $i$, as well as the elements $\overline{C_{ij}}\equiv
\langle(P_i-\overline{P_i})(P_j-\overline{P_j})\rangle$ of the new
$5\times 5$ covariance matrix.  $\Delta\chi^2$ is then computed
between pairs of cosmological models, analogous to
Eq.~(\ref{eq:meanchi2}) for the peak counts.

In order to find the information beyond the power spectrum, we first
calculate the $\Delta\chi^2$ using the power spectrum alone. We then
combine the five peak counts and the five power spectra into a vector
of 10 observables, and compute the cross--terms,
$\overline{C_{ij}}\equiv
\langle(P_i-\overline{P_i})(N_j-\overline{N_j})\rangle$, to obtain the
elements in the off-diagonal blocks of the $10\times10$ covariance
matrix.  This allows us to calculate $\Delta\chi^2$ from the
$N(\kappa)+P(\ell)$ combination, taking into account their
correlations.

\begin{table}
\begin{tabular}{|c|l|cc|} 
\hline
observable type & cosmology & noiseless $\Delta\chi^2$ & noisy $\Delta\chi^2$ \\
 \hline
 Peak Counts ($\Delta\chi^2_{N}$)     & Fiducial & 5.16 & 5.89 \\
 Power Spectrum  ($\Delta\chi^2_{P}$) & and & 17.06 & 8.12 \\
 Combination ($\Delta\chi^2_{NP}$)    & High-$\sigma_8$ & 37.07 & 16.36 \\
$\Delta\chi^2_{NP}/(\Delta\chi^2_{N}+\Delta\chi^2_{P})$ &  & 1.67 & 1.17 \\
 \hline
 Peak Counts ($\Delta\chi^2_{N}$)     & Fiducial & 5.01&  5.09 \\
 Power Spectrum  ($\Delta\chi^2_{P}$) & and & 13.03 & 5.76 \\
 Combination ($\Delta\chi^2_{NP}$)    & Low-$\sigma_8$ & 26.79 & 11.87 \\
$\Delta\chi^2_{NP}/(\Delta\chi^2_{N}+\Delta\chi^2_{P})$ &  & 1.49 & 1.09 \\
 \hline
 Peak Counts ($\Delta\chi^2_{N}$)     & Fiducial & 3.61 & 4.02  \\
 Power Spectrum  ($\Delta\chi^2_{P}$) & and & 17.69 & 6.15 \\
 Combination ($\Delta\chi^2_{NP}$)    & High-$\Omega_{m}$ & 32.42 & 11.65 \\
$\Delta\chi^2_{NP}/(\Delta\chi^2_{N}+\Delta\chi^2_{P})$ &  & 1.52 & 1.15 \\
 \hline
 Peak Counts ($\Delta\chi^2_{N}$)     & Fiducial & 4.39  & 4.44 \\
 Power Spectrum  ($\Delta\chi^2_{P}$) & and & 16.49 & 5.61 \\
 Combination ($\Delta\chi^2_{NP}$)    & Low-$\Omega_{m}$ & 29.47 & 10.94 \\
$\Delta\chi^2_{NP}/(\Delta\chi^2_{N}+\Delta\chi^2_{P})$ &  & 1.41 & 1.09 \\
 \hline
 Peak Counts ($\Delta\chi^2_{N}$)     & Fiducial & 0.98 & 0.65  \\
 Power Spectrum  ($\Delta\chi^2_{P}$) & and & 0.92 & 0.29 \\
 Combination ($\Delta\chi^2_{NP}$)    & High-$w$ & 2.79 & 0.84 \\
$\Delta\chi^2_{NP}/(\Delta\chi^2_{N}+\Delta\chi^2_{P})$ &  & 1.46 & 0.90\\
 \hline
 Peak Counts ($\Delta\chi^2_{N}$)     & Fiducial & 0.44 & 0.36 \\
 Power Spectrum  ($\Delta\chi^2_{P}$) & and & 0.64 & 0.19\\
 Combination ($\Delta\chi^2_{NP}$)    & Low-$w$ & 1.69 & 0.51 \\
$\Delta\chi^2_{NP}/(\Delta\chi^2_{N}+\Delta\chi^2_{P})$ & & 1.57 & 0.92 \\
 \hline
\end{tabular}
\caption[]{\textit{$\Delta\chi^2$ from peak counts, power spectra, and
their combination, computed between the fiducial model and six other
models varying $\sigma_8$, $w$, and $\Omega_m$
independently. $\Delta\chi^2_{NP}$ denotes the $\Delta\chi^2$ from the
combination of peak counts and power spectrum, including their
correlations; $\Delta\chi^2_{N}$ and $\Delta\chi^2_{P}$ denote the
individual $\Delta\chi^2$'s.  1,000 noise--free or noisy maps are used
for each of the model.  Source galaxies are at $z_s=2$.}}
\label{tab:chisqSimuPK_PS_PKPS}
\end{table}

The results are shown in Table~\ref{tab:chisqSimuPK_PS_PKPS}.
Comparing the individual $\Delta\chi^2$'s first, we see that in the
noisy maps, the sensitivity of peak counts is roughly comparable to
the power spectrum, although about $\sim50\%$ weaker for $\sigma_8$
and $\Omega_m$, and about twice stronger for $w$.  In the noiseless
maps, the power spectrum is more sensitive, especially for $\sigma_8$
and $\Omega_m$.  This shows, interestingly, that the power spectrum
sensitivity is much more degraded by noise than the peak counts. This
is not surprising, given that the constraints from the power spectrum
are dominated by linear fluctuations on relatively large
scales~\cite{FH07, Huterer02}, with noise adding linearly to the
large-scale structure signal.  In other words, unlike for peak counts,
adding noise does not change the signal (i.e., the difference
$\Delta\overline{P_i}$ between two cosmologies), as long as the galaxy
noise is independent of cosmology, whereas the noise increases the
variances $\langle(P_i-\overline{P_i})(P_j-\overline{P_j})\rangle$.

Inspecting next the combined $\Delta\chi^2$'s (shown in the third row
in each section of Table~\ref{tab:chisqSimuPK_PS_PKPS}), we find that
these are comparable to adding the two individual $\Delta\chi^2$
values. This would be expected if there were no cross--correlations
between power spectra and peak counts. Indeed, this result appears
consistent with the negligible correlation between the 3d space
density of clusters and the 2d convergence power
spectrum~\cite{FH07,TB07}.  Interestingly, however, in the noiseless
maps, the $N(\kappa) + P(\ell)$ combination yields a {\em better}
sensitivity than adding two observables independently (by $\sim50\%$;
see each 4th row in the Table).  This ``sum greater than its parts''
effect can arise whenever $N(\kappa)$ and $P(\ell)$ have a nonzero
correlation $\langle \Delta N \Delta P\rangle \neq 0$, and the
cosmology-induced changes $\delta N$ and $\delta P$ do not obey the
same correlation.  It can be verified, after some algebra, that in the
case of two observables $N$ and $P$, individually yielding
$\Delta\chi^2_N$ and $\Delta\chi^2_P$, the condition for
$\Delta\chi^2_{NP} > \Delta\chi^2_N+\Delta\chi^2_P$ is $ (\delta N
\delta P) / \langle \Delta N \Delta P\rangle <
(\Delta\chi^2_N+\Delta\chi^2_P)/2$.  By inspecting the cross-terms in
our $10\times10$ covariance matrix, we have verified that this
condition is satisfied for each $N_i$ and $P_j$ pair whose combination
enhances their $\Delta\chi^2$.  For example, when $\sigma_8$ is
decreased, the peak counts in the lowest bin ($N_1$) decrease, as do
the power spectra -- however, the covariance matrix predicts an
anti-correlation between $N_1$ and all five $P_j$'s.

\subsection{Redshift Tomography}
\label{subsec:Tomog}

\begin{table}
\begin{tabular}{|c|l|cccc|} 
\hline
source & cosmology & \multicolumn{2}{c}{noiseless $\Delta\chi^2$} & \multicolumn{2}{c}{noisy $\Delta\chi^2$}  \vline\\
 & & unscaled & scaled & unscaled & scaled \\
 \hline
 z2 & Fiducial & 5.16 & 0.46 & 5.89 & 4.29 \\
 z1 & and & 3.36 & 0.66 & 2.67 & 2.56 \\
 z12 & High-$\sigma_8$ & 5.99 & 0.91 & 6.16 & 4.84 \\
 z12/(z2+z1) &  & 0.70 & 0.81 & 0.72 & 0.71 \\
 \hline
 z2 & Fiducial & 5.01 & 0.34 & 5.09 & 3.67 \\
 z1 & and & 3.27 & 0.73 & 2.23 & 2.23 \\
 z12 &Low-$\sigma_8$ & 5.90 & 0.94 & 5.29 & 4.05 \\
 z12/(z2+z1) &  & 0.71 & 0.88 & 0.72 & 0.69 \\
 \hline
 z2 & Fiducial & 3.61 & 0.033 & 4.02 & 2.46 \\
 z1 & and & 4.47 & 0.044 & 2.97 & 2.15 \\
 z12 &High-$\Omega_{m}$ & 5.41 & 0.067 & 4.51 & 3.12 \\
 z12/(z2+z1) &  & 0.67 & 0.87 & 0.65 & 0.68 \\
 \hline
 z2 & Fiducial & 4.39 & 0.053 & 4.44 & 2.56 \\
 z1 & and & 5.30 & 0.051 & 2.86 & 2.23 \\
 z12 &Low-$\Omega_{m}$ & 6.51 & 0.082 & 4.76 & 3.15 \\
 z12/(z2+z1) &  & 0.67 & 0.79 & 0.65 & 0.66 \\
 \hline
 z2 & Fiducial & 0.98 & 0.47 & 0.65 & 0.27 \\
 z1 &and & 1.24 & 0.58 & 0.40 & 0.20 \\
 z12 &High-$w$ & 1.57 & 0.83 & 0.70 & 0.37 \\
 z12/(z2+z1) &  & 0.71 & 0.80 & 0.67 & 0.79 \\
 \hline
 z2 & Fiducial & 0.44 & 0.27 & 0.36 & 0.16 \\
 z1 & and & 0.94 & 0.39 & 0.37 & 0.19 \\
 z12 &  Low-$w$ & 1.12 & 0.56 & 0.48 & 0.28 \\
 z12/(z2+z1) & & 0.81 & 0.85 & 0.66 & 0.80 \\
 \hline
\end{tabular}
\caption[]{\textit{This table examines a simple case of tomography
with two redshifts.  $\Delta\chi^2$ values are shown between the
fiducial model and six other models varying $\sigma_8$, $w$, and
$\Omega_m$, for both unscaled and scaled peak height distributions,
obtained using 1,000 noise--free or noisy maps.  Source galaxies are
located at $z_s=1$, at $z_s=2$, or at both redshifts (denoted by z1,
z2, and z12). The rows labeled by ``z12/(z2+z1)'' show the combined
$\Delta\chi^2$ divided by the sum of the individual $\Delta\chi^2$ of
z1 and z2.}}
\label{tab:chisqSimuZ2Z1Z12}
\end{table}

Our analysis above relied on a single source galaxy redshift at
$z_s=2$.  In a realistic survey, there will of course be a
distribution of galaxy redshifts. Using galaxies at different
redshifts (``tomography'') could, in principle, strengthen
cosmological constraints significantly, despite the strong
correlations in the signal measured at different source galaxy planes
~\cite{Hu99, Huterer02}.

Here we evaluate the benefits of tomography in the simplest case of
having source galaxies at two distinct redshifts. We calculate the
$\Delta\chi^2$ from the peak counts, as before, from source galaxies
($15$ ${\rm arcmin^{-2}}$) separately at $z_s=1$ and $z_s=2$, using
five convergence bins at each redshift.  We then combine these, and
calculate the $\Delta\chi^2$ using both redshifts (i.e. a total of
$30$ galaxies ${\rm arcmin^{-2}}$) and their joint $10\times10$
covariance matrix.  This calculation includes the covariance across
the two redshift bins, and is analogous to the combination of the peak
counts and the power spectra described in the previous section.

The results are shown in Table~\ref{tab:chisqSimuZ2Z1Z12}.  In each
section of the table, the first three rows show $\Delta\chi^2$ at
$z_s=2$, $z_s=1$, and the combined constraints.  The fourth row shows
the ratio of the combined $\Delta\chi^2$ to the sum of the individual
$\Delta\chi^2$'s at the two redshifts. This last quantity checks the
importance of the covariance between the two redshifts. It would be
unity if the two PDFs were completely independent, but can be either
larger or smaller than unity if the correlations between the two
redshifts are important (as discussed in previous section).

Comparing the individual redshifts first, as the first two rows in the
table show, in the noisy maps, $z_s=2$ generally yields a better
sensitivity than $z_s=1$. The only exception is the low-$w$ case, when
the sensitivities at the two redshifts are comparable (with $z_s=1$
only slightly better). This is consistent with our results in Paper I,
in which we have also found that the sensitivity to a combination of
($\sigma_8.w$) increases with source galaxy redshift.  The advantage
of higher redshift is explained by the accumulation of a larger
overall lensing signal, when going to a large distance.  In our
noiseless fiducial maps, we have $\sigma_\kappa=0.022$ and 0.013 at
$z_s=2$ and $z_s=1$, respectively, which is to be compared to our
assumed noise of $\sigma_{\rm noise}=0.023$ and $\sigma_{\rm
noise}=0.019$ at $z_s=2$ and $z_s=1$.  Comparing the noisy and the
noiseless results in Table~\ref{tab:chisqSimuZ2Z1Z12}, we see that
adding noise to the unscaled maps for $z_s=1$ always hurts, and
decreases the $\Delta\chi^2$ values. In contrast, adding noise {\em
increases} the $\Delta\chi^2$ values for $z_s=2$.  We find that at
both redshifts, adding noise enhances the difference in the total
number of peaks (this counterintuitive result is explained in detail
in the next section).  However the $\Delta\chi^2$ depends not only on
the total number of peaks, but also on the shape of the peak height
distribution.  At $z_s=1$, where $\sigma_\kappa$ is well below the
noise $\sigma_{\rm noise}$, the peak count shape distribution is much
more vulnerable to the noise.

It is worth noting that, apart from the importance of noise, there are
trends with redshifts arising from the cosmological dependence of (i)
geometrical distance factors in the lensing kernel, and (ii) from the
growth of the matter perturbations.  Individually, both of these
depend on cosmology, with the induced differences increasing with
redshift, and strengthening the sensitivity.  However, there are
cancellations when the effects from the geometry and growth work in
the opposite direction, which weakens the overall sensitivity. This
cancellation can be worse at high redshift. This explains why in the
noiseless case, $z_s=1$ is, in fact, better than $z_s=2$, for both $w$
and $\Omega_m$.  Reducing $\Omega_m$ or increasing $w$ both result in
flatter growth (i.e. larger density fluctuations at high redshift, for
fixed $\sigma_8$), which is canceled by a reduction in the lensing
kernel.  For $\sigma_8$, when only growth effects are present, and
there are no such cancellations, the sensitivity always increases with
redshift.

Inspecting next the combined $\Delta\chi^2$'s, we see that in general,
tomography does {\em not} significantly improve the sensitivity,
compared to having only the more sensitive of the two redshifts.  This
is partly due to the fact that the less sensitive of the two redshifts
is significantly less sensitive, and partly due to the covariance
between the two redshifts, which reduces the combined $\Delta\chi^2$
by $\sim$30\% compared to having two uncorrelated measurements. (The
typical value of the ``covariance parameter'' shown in the fourth row
in each section of Table~\ref{tab:chisqSimuZ2Z1Z12} is $\sim$0.7.).
We emphasize, however, that the change in the peak counts induced by
each parameter has, in general, a different redshift-dependence.
Therefore, tomography can still be very useful to improve the
marginalized constraints, whenever there is a strong degeneracy
between parameters at a single redshift (see discussion of
marginalized constraints below).

\subsection{Why Does Noise Increase the Signal-to-Noise?}
\label{subsec:noise}

An interesting finding in this paper is that adding random noise can
sometimes boost the cosmology sensitivity of peak counts (i.e., at
high redshift, as mentioned in last section).  A similar result -
namely that the difference in the total number of peaks is increased
when noise is added - was found (but not explored) in Paper I.  This
is a counter-intuitive result that we investigate here.

For simplicity, the discussion below will be restricted to the
$\Delta\chi^2$ obtained from the total number of peaks (effectively
using a single convergence bin).  We find that noise boosts these
$\Delta\chi^2$, as well.  For example, the mean number of peaks in our
fiducial, high-$\sigma_8$ and high-$\Omega_m$ models are $2337.9$,
$2326.4$ and 2339.3 in noise free maps, and $3414.6$, $3362.2$ and
$3369.7$ in noisy maps, respectively. The r.m.s of the total number of
peaks in the fiducial model is 35 in the noise--free maps and is only
slightly larger, 38, in the noisy maps. This implies that, for
example, for $\sigma_8$, noise increases $\Delta\chi^2$ from
$11.5/35\approx 0.3$ to $52.4/38\approx 1.4$.

Below, we will use the predictions in a GRF to explain such an
increase.  The advantage of using a GRF is that the peak counts are
analytically predictable, allowing us to understand the effect of the
noise exactly.  Also, as we showed earlier, the cosmology-induced
differences in the peak counts are generally close to those in a GRF
(even though the peak height distributions are dissimilar). Therefore,
it is reasonable to use the GRF as a guide to understand a boost in
the $\Delta\chi^2$.

The galaxy shape noise added to our maps is assumed to be uncorrelated
in each pixel -- this corresponds to a GRF with a flat power spectrum,
or ``white noise''.  Applying the definition of $\sigma_p$ in
Eq.~(\ref{eq:GRFeq5}) to such white noise, and assuming a smoothing
scale $\theta_G$, we find the following relations between
$\sigma_{n0}$, $\sigma_{n1}$, $\sigma_{n2}$:
\BA\label{eq: noise add distinct pow 1}
\sigma^2_{n2}&=&\sigma^2_{n1}\frac{4}{\theta^2_G}\\&=&\sigma^2_{n0}\frac{8}{\theta^4_G},
\EA
where $\sigma_{n0}$ is given by Eq.~(\ref{eq:ellipnoise2}). In our
case, with $n_{\rm gal}=15~{\rm arcmin^{-2}}$, and 1 arcmin smoothing
at $z_s=2$, we find the numerical values
$(\sigma^2_{n0},\sigma^2_{n1},\sigma^2_{n2})=(0.00051354,3.6975~{\rm
deg^{-2}}, 53244~{\rm deg^{-4}})$.

For arbitrary power spectra, the total number of peaks is given by a
constant $\times (\sigma^2_2/\sigma^2_1)$ (see
Eq.~(\ref{eq:GRFeq6})). In the following analysis, we drop this
constant for convenience.  Let us next denote the $\sigma$'s in the
first cosmology by $\sigma_{0}$, $\sigma_{1}$, $\sigma_{2}$, and in
the second cosmology by $\sigma^\prime_{0}$, $\sigma^\prime_{1}$,
$\sigma^\prime_{2}$.  In the absence of noise, the difference in total
number of peaks is given by
\BA
\label{eq: noise add distinct pow 2} 
\Delta n_{\rm pk}&=&\frac{\sigma^2_2}{\sigma^2_1}-\frac{\sigma^{\prime2}_2}{\sigma^{\prime2}_1}\\
&=&\frac{\sigma^2_2}{\sigma^2_1}(1-\frac{r_2}{r_1})
\EA
where $r_1 \equiv \sigma^{\prime2}_1/\sigma^{2}_1$ and $r_2 \equiv
\sigma^{\prime2}_2/\sigma^{2}_2$.  Since the noise is assumed to be
uncorrelated with the noise-free convergence field, the $\sigma$'s of
the of the noise-free field and of the noise field add
linearly. Therefore, the difference in the total number of peaks,
after the noise is added, is given by:
\BA
\label{eq: noise add distinct pow 3}
\Delta n_{\rm pk,noise}&=&\frac{\sigma^2_2+\sigma^2_{n2}}{\sigma^2_1+\sigma^2_{n1}}-\frac{\sigma^{\prime2}_2+\sigma^{2}_{n2}}{\sigma^{\prime2}_1+\sigma^{2}_{n1}}\\
&=&\frac{\sigma^2_2}{\sigma^2_1}\frac{(1+a_2)}{(1+a_1)}-\frac{\sigma^2_2}{\sigma^2_1}\frac{(r_2+a_2)}{(r_1+a_1)}\\ 
\label{eq:noisycounts}
&=&\frac{\sigma^2_2}{\sigma^2_1}\left[\frac{(1+a_2)}{(1+a_1)}-\frac{(r_2+a_2)}{(r_1+a_1)}\right],
\EA
where $a_1\equiv\sigma^{2}_{n1}/\sigma^{2}_1$ and $a_2
\equiv\sigma^{2}_{n2}/\sigma^{2}_2$.

We now look at the magnitudes of $a_1$ and $a_2$ (which express the
importance of noise relative to the cosmological lensing signal) and
$r_1$ and $r_2$ (which express the changes caused by the
cosmology). Considering, as an example, the fiducial model and the
high-$\sigma_8$ model as the second (primed) cosmology, we have $a_1 =
3.368$, $a_2 = 5.537$, $r_1 = 1.197$ and $r_2 = 1.189$.  Clearly,
$r_1$ and $r_2$ are very close to each other, whereas $a_2$ differs
significantly from $a_1$.  Looking at Eqs.~(\ref{eq: noise add
distinct pow 3}) and (\ref{eq: noise add distinct pow 2}), we see that
$r_2/r_1\approx 1$ implies $\Delta n_{\rm pk}$ will be small, and
comparing the factors multiplying the term $\sigma_2^2/\sigma_1^2$, we
infer $\Delta n_{\rm pk,noise}>\Delta n_{\rm pk}$ as long as $r_1
\approx r_2 > 1$ and $a_2>a_1>1$.  In our case, we find $\Delta n_{\rm
pk,noise}=54.2$ and $\Delta n_{\rm pk}=15.3$.  This then implies a
significant increase in the total $\Delta\chi^2$, provided that the
r.m.s of total number of peaks doesn't increase much (which is indeed
the case; we find that the r.m.s. increases by $\approx8\%$).  The
first of the two conditions responsible for $\Delta n_{\rm
pk,noise}>\Delta n_{\rm pk}$, namely that $r_1\approx r_2$, says that
the first and second derivatives of the correlation function
(Eq.~\ref{eq:GRFeq5}) scale very similarly with our parameter,
$\sigma_8$.  This makes sense, and would indeed hold strictly (with
$r_1=r_2$) in the linear regime. However, although we are using GRFs,
we adopt the nonlinear power spectra from the simulations, and
therefore the scaling with $\sigma_8$ is stronger than linear on small
scales.  We have verified that the small ($<1\%$) difference we find
between $r_1$ and $r_2$ is not a numerical artifact, and a similar
difference is present when we compute the $\sigma$'s from the
theoretical power spectra~\cite{Smith+03}.  The second of the two
conditions is that $a_2$ differs significantly from $a_1$, with
$a_2>a_1$ (note that since noise is added on top of the maps,
$(a_2,a_1)>1$ always holds), but not overwhelmingly $> 1$ (otherwise
noise would dominate the lensing signal, and there would be no
distinction).  This also makes sense: the relative importance of the
noise and the cosmological lensing signal is wavelength-dependent.
More specifically, the latter decreases with increasing wavenumber,
and therefore noise is increasingly important on small scales -- as a
result, whenever the noise is significant, it has a bigger effect on
the second derivatives than on the first.

The above result raises two more questions. First, what is the ideal
noise level, which maximizes the signal $\Delta n_{\rm pk}$ (or,
ultimately, the actual $\Delta\chi^2$ values in the simulated WL
maps)?  Also, how much noise is too much? As the noise is increased,
eventually it must hurt, and reduce $\Delta n_{\rm pk}$ below its
noiseless value. At what level of noise does this occur?  In order to
answer these questions, we first repeated the analysis in the GRF
case, but multiplied the noise $\sigma_{n0}$ by a constant factor.
This increases $\sigma_{n1}$ and $\sigma_{n2}$ by the same factor, and
$a_1$ and $a_2$ by the square of this factor, so the dependence of
$\Delta n_{\rm pk}$ on the noise level can be simply obtained from
equation~(\ref{eq:noisycounts}).  In practice, we went through the
exercise of adding random noise with different amplitudes to the mock
GRF maps. We found that the difference in the peak counts (between the
fiducial and the high-$\sigma_8$ models) followed very accurately the
predictions from equation~(\ref{eq:noisycounts}).  Having the maps
then allowed us to compute the variance in the number of peaks
$\langle\delta n_{\rm pk}^2\rangle$ (in our fiducial model).  These
results are shown in the 2nd and 3rd columns of
Table~\ref{tab:noiselevels}.  We then performed the same exercise for
the simulated WL maps (again between the fiducial and the
high-$\sigma_8$ models), adding different levels of noise, and
recomputing $\Delta n_{\rm pk}$ (shown in the 4th column of
Table~\ref{tab:noiselevels}), as well as the $\Delta\chi^2$ both
scaled and unscaled, as defined above (5th and 6th
columns).\footnote{In this last analysis, we used fixed $\kappa$ bins
with roughly equal counts.  This was necessary to avoid choosing
different boundaries, for each noise level, by the ad-hoc optimization
procedure used above.  This causes $\Delta\chi^2$ values in
Table~\ref{tab:noiselevels} to differ slightly from those in
Table~\ref{tab:chisq Simu GRF Z2} but should not affect our argument
and conclusions here.  Bin boundaries are discussed in detail in
\S~\ref{subsec:usebin} below.}

Table~\ref{tab:noiselevels} shows that there is an "ideal" noise
level, at which $\Delta n_{\rm pk}$ is maximized.  This turns out to
be approximately half the noise we adopted.  There is also a level
(approximately twice larger than we adopted), beyond which noise
actually hurts in the absolute sense, i.e. $\Delta n_{\rm pk}$ becomes
smaller than in the noise-free case.  As the noise is increased
further, $\Delta n_{\rm pk}$ tends to zero, as it should.
Interestingly, these conclusions hold, both in the GRF and the
simulated maps (2nd and 4th rows).  We found that the variance in the
peak counts does not change significantly as noise is added (either
for a GRF or in our simulated maps; the GRF case is shown in the 3rd
column).  Most importantly, the actual $\Delta\chi^2$ values are also
maximized at $\sim$half of our original noise (5th and 6th columns);
the unscaled $\Delta\chi^2$, however, drops quickly below the
noiseless case when the noise exceeds the original value.

The above analysis demonstrates that the naive intuition, namely that
noise can only decrease the signal-to-noise ratio (which is manifestly
true when the signal and noise add linearly), no longer holds in our
case.  This naive intuition is known to fail when the "signal" is a
nonlinear function of the noise.  Indeed, noise can amplify the signal
non-linearly, via a phenomenon called "stochastic resonance"
\cite{Gammaitoni+98}, under three generic conditions: (i) the presence
of some form of threshold in the definition of the signal, (ii) a weak
coherent input, and (iii) a source of noise that adds to the coherent
input.  It is interesting to note that these three conditions are
satisfied in our WL peak-counts, and hence WL peaks appear to be an
example of this phenomenon; this connection is worth exploring further
in future work.

\begin{table}
\begin{tabular}{|l|c|c|c|c|c|} 
\hline
 Noise  & \multicolumn{2}{c}{GRF}\vline & \multicolumn{3}{c}{Simulations}\vline \\
 Level  & $\Delta n_{\rm pk}$ & $\langle\delta n_{\rm pk}^2\rangle$ & $\Delta n_{\rm pk}$ & $\Delta\chi^2$ & $\Delta\chi^2$  \\
        &                     &                                     &                     &  (scaled)      &  (unscaled)  \\
\hline
0	&  15.3  &  35.0   &   11.5  &  7.20	& 0.43\\
0.5	&  75.2  &  42.4   &   65.5  &  8.41	& 5.08\\
1$\star$&  54.2  &  37.7   &   52.4  &  6.57	& 4.23\\
2	&  20.1  &  31.5   &   21.1  &  2.16 	& 1.00\\
4	&   5.6  &  31.1   &    5.8  &  0.27	& 0.10\\
\hline
\end{tabular}
\caption[]{\textit{The difference in the total number of peaks between
    the fiducial model and the high-$\sigma_8$ model, as a function of
    the level of the noise, in the GRF case (2nd column) and in the
    simulated WL maps (4th column).  The first column shows the
    numerical factor by which we multiplied the original noise level
    (the third row, marked with a star, corresponds to the original
    noise).  The 3rd column shows the r.m.s. of the peak counts in the
    fiducial model and the 5th and 6th columns show $\Delta\chi^2$
    values as a function of the noise.  In all cases, we find that the
    two cosmologies are best distinguished when approximately half of
    our original noise is added to the maps; the distinction rapidly
    decreases for noise $\gsim$ twice our original value.}}
\label{tab:noiselevels}
\end{table}

Finally, when we compare the fiducial model to the high-$\Omega_m$
cosmology, we find that noise boosts the signal even more
significantly than for the high-$\sigma_8$ model.  In this case, we
have $a_1 = 3.368$, $a_2 = 5.537$, $r_1 = 1.1707$, and $r_2 = 1.1716$.
Clearly, both conditions above are still satisfied, and we obtain
$\Delta n_{\rm pk,noise}=43.6$ compared to $\Delta n_{\rm pk}=-1.7$.
Note that in this case, $\Delta n_{\rm pk}$ is negative, because
$r_1<r_2$.  For reference, the total number of peaks in our fiducial,
high-$\sigma_8$ and high-$\Omega_m$ models, calculated through
Eq.~(\ref{eq:GRFeq6}), is 2275.0, 2259.7 and 2276.7 in noise free
maps, and 3431.3, 3377.1 and 3387.7 in noisy maps.  These differ from
the total counts in the simulations, quoted in the beginning of this
section, by $\Delta n \approx -60$ and $\Delta n \approx +20$ in the
noise-free and noisy cases, respectively.  However, as already noted
above, the cosmology-induced differences are very similar in the
simulations and the GRF case, as summarized in
Table~\ref{tab:dn-noise}.

\begin{table}
\begin{tabular}{|l|c|c|c|c|} 
\hline
 $\Delta n_{\rm pk}$ & \multicolumn{2}{c}{(fiducial)-(high-$\sigma_8$)}\vline & \multicolumn{2}{c}{(fiducial)-(high-$\Omega_m$)}\vline \\
            & noisy     & noiseless         & noisy     & noiseless \\
\hline
GRF         & 54.2      & 15.3               & 43.6      & -1.7\\
Sim         & 52.4      & 11.5               & 44.9      & -1.4\\
\hline
\end{tabular}
\caption[]{\textit{The change in the total number of peaks when the
    fiducial model is compared to the high-$\sigma_8$ or the
    high-$\Omega_m$ model.  As the table shows, the changes are
    significantly enhanced by noise, and are similar in the GRF and
    the simulations.}}
\label{tab:dn-noise}
\end{table}

\subsection{Can $\Delta\chi^2$ be Interpreted as a Likelihood?}
\label{subsec:MeaningDChi2}

\begin{figure}[htp]
\centering
\includegraphics[width=8 cm]{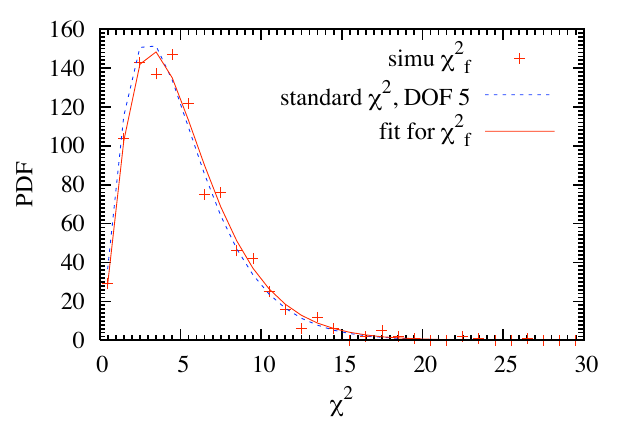} \hfill
\caption[]{\textit{The PDF of $\chi^2_{f}(r)$ over different
    realizations within our fiducial model (red crosses).  The data is
    fit well by true chi-squared distributions: the solid red and blue
    dashed curves show true $\chi^2$-distributions with 5.23 and 5
    degrees of freedom.  Noise-free, unscaled maps were used for this
    figure, with source galaxy redshift $z_s=2$.}}
\label{fig:noncen_chisq_fit}
\end{figure}

So far, we have quoted the $\Delta\chi^2$ values based on differences
in mean peak counts between models.  An important question is whether
these $\Delta\chi^2$ can be interpreted as likelihoods, or confidence
levels on parameter estimates.  If the observables (in our case, the
peak counts in each bin) were Gaussian distributed, and if they
depended linearly on the the parameters (in our case, the cosmological
parameters), then our $\Delta\chi^2$ would follow true $\chi^2$
distributions.  When fitting a single parameter, as in our case above,
$\Delta\chi^2=1, 4, 9$ would then correspond to the usual 68.3\%,
95.4\%, 99.7\% confidence levels.\footnote{Note that we neither have
an actual data set, nor do we perform a $\chi^2$ minimization to find
the best-fit parameters. We are thus effectively assuming that the
mean peak counts in our fiducial model are the data, yielding our
fiducial parameter as the best fit; we can then find the confidence
limits corresponding to the other six models.}

To see how good the above approximations are, in
Fig.~\ref{fig:noncen_chisq_fit}, we first show the distribution of
$\chi^2_{f}(r)$ over different realizations in the fiducial model
itself, shown by the red crosses (computed from Eq.~(\ref{eq:chisq1}),
using noiseless, unscaled maps, and $z_s=2$).  We fit these data with
a standard chi-squared distribution $P_{\chi^2}(a\chi^2, DOF)$,
treating a linear scaling constant $a$ and the number of degrees of
freedom DOF as free parameters.  We find best-fit values of $a=1.0015$
and $DOF = 5.23$.  Reassuringly, the fit, shown by the red solid
curve, is very good, with $a$ close to 1, and $DOF$ close to 5, the
number of bins we used.  For comparison, $P_{\chi^2}$ with $a = 1$,
$DOF = 5$ is also shown as the blue dashed curve.  Clearly,
$\chi^2_{f}(r)$ closely follows a chi-squared distribution expected if
the deviation of peak counts from the mean were Gaussian.

These results justify interpreting our $\Delta\chi^2$'s in the Tables
above as (single-parameter) confidence levels.  As seen in
Table~\ref{tab:chisq Simu GRF Z2}, the unscaled, noisy maps when
$\sigma_8$ and $\Omega_m$ are varied correspond to ``$2-2.5\sigma$''
differences from the fiducial model; $w$ variations correspond to
``$0.5-0.8\sigma$'' differences.

\subsection{Impact of the Choice of Binning}
\label{subsec:usebin}

In all our results above, we have used a fixed number of (five) bins,
and performed only an ad-hoc optimization of the bin boundaries by
hand.  It is important to ask how our results are affected both by the
number of bins, and by the placement of the bin boundaries. Ideally,
one could use arbitrarily fine binning, and avoid such questions; in
practice, we are limited by the finite number of realizations we can
simulate.

\begin{figure}[htp]
\centering
\includegraphics[width=8 cm]{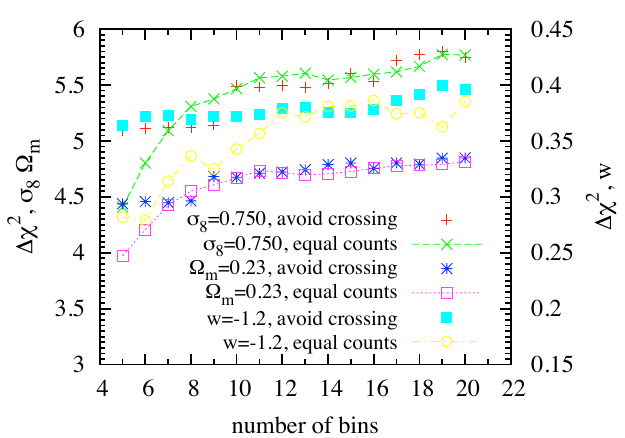} \hfill
\caption[]{\textit{$\Delta\chi^2$ from peak counts, as a function of
    the number of convergence bins. The fiducial model was compared to
    the low-$\sigma_8$, low-$\Omega_m$, and low-$w$ models, using
    noisy unscaled maps with $z_s=2$.  Bin boundaries were chosen such
    that each bin contains equal counts, or such that ``crossings'' of
    the peak-count PDFs in a pair of cosmologies are avoided within
    bins (as labeled).  The y-axis labels on the left refer to the
    $\sigma_8$ and $\Omega_m$ cases; the labels on the right to the
    $w$ case.  }}
\label{fig:chisq_binnum}
\end{figure}

In Fig.~\ref{fig:chisq_binnum}, we show the $\Delta\chi^2$ in noisy
unscaled maps, with $z_s=2$, between the fiducial model and the
low-$\sigma_8$, low-$w$ and low-$\Omega_m$ models, as a function of
the number of bins. We chose the bin boundaries either by following
the approach of avoiding ``crossings'' of the peak-count PDFs in a
pair of cosmologies within bins, or such that the mean number of peaks
in each bin were the same.  As the figure shows, the ``avoid
crossings'' approach works quite well, and from 5 to 20 bins, the
$\Delta\chi^2$'s increase only modestly (by $\approx 10\%$).  The
``equal counts'' approach does more poorly (yielding smaller
$\Delta\chi^2$) when the number of bins is small, but converges to a
very similar values once the number of bins is $\gsim 15$.  These
results give reassurance that we have a sufficient number of bins and
the $\Delta\chi^2$'s shown in the Tables above have converged to
within $\sim 10\%$.

\begin{figure}[htp]
\centering
\includegraphics[width=8 cm]{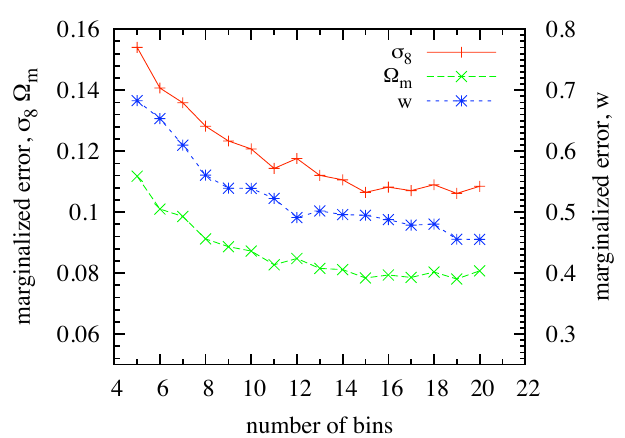} \hfill
\caption[]{\textit{The marginalized errors of the three cosmological
    parameters $\sigma_8$, $\Omega_m$, $w$ from the combination of the
    peak counts and power spectrum, as a function of the number of
    convergence bins. The fiducial model was compared to the
    high/low-$\sigma_8$, high/low-$\Omega_m$, and high/low-$w$ models,
    using noisy unscaled maps with $z_s=2$.  Bin boundaries were
    chosen such that each bin contains equal counts for peak counts
    and equally spaced with the cut at $\ell=20,000$ for the power
    spectrum.  The y-axis labels on the left refer to the $\sigma_8$
    and $\Omega_m$ cases; the labels on the right to the $w$ case.  }}
\label{fig:merr_binnum}
\end{figure}

In the next section, we vary multiple parameters simultaneously.  In
Figure~\ref{fig:merr_binnum}, we show the marginalized errors of the
three cosmological parameters $\sigma_8$, $\Omega_m$, $w$ from the
combination of the peak counts and the power spectrum, as a function
of the number of bins (in noisy unscaled maps, with $z_s=2$). When
choosing the boundaries, we applied the ``equal counts" approach for
the peaks and the "equally spaced" approach for the power
spectrum. The details of computing the marginalized error is explained
in the next section. The figure shows that the five bins are not
sufficient in this case; however, the marginalized error converges to
within $< 10\%$ once the number of bins is $\gsim 15$.

\subsection{Forecasting Marginalized Errors}
\label{subsec:Marginalized Error}

\begin{table}
\begin{tabular}{|c|c|c|c|} 
\hline
marginalized error & $\sigma_8$ & $w$ & $\Omega_m$\\
 \hline
 z2 & 0.0065 & 0.030 & 0.0057 \\
 z1 & 0.0078 & 0.036 & 0.0057 \\
 z2+z1 & 0.0024 & 0.018 & 0.0022 \\
 Power Spectrum ($z_s=2$) & 0.0047 & 0.026 & 0.0028 \\
 \hline
 z2+Power Spectrum &  0.0026 & 0.012 & 0.0019 \\
 z1+Power Spectrum &  0.0037 & 0.020 & 0.0026 \\
 tomography combined &  0.0012 & 0.0096 & 0.0010 \\
 combined/( z2+Power Spectrum) & 0.47 & 0.79 & 0.52 \\
 \hline
\end{tabular}
\caption[]{\textit{Marginalized 68\% errors, in our noisy maps, on the
    cosmological parameters $\sigma_8$, $w$, and $\Omega_m$.  In the 
    top half of the table, peak counts and power spectra are considered
    separately.  From top to bottom:
(i)    counts alone at $z_s=2$;
(ii)   counts alone at $z_s=1$;
(iii)  counts alone with both $z_s=1$ and $z_s=2$;
(iv)   power spectrum alone at $z_s=2$.
    In the bottom half of the table, counts and the power spectrum are
    combined.  From top to bottom:
(v)    combining counts and power spectrum at $z_s=2$;
(vi)   combining counts and power spectrum at $z_s=1$;
(vii)  combining the above two cases to use all 4 observables -- peak counts 
      and power spectrum at $z_s=2$ and $z_s=1$;
and finally (viii) the last combined results (row vii) divided by the
    ``z2+Power Spectrum" results (row v).  
Each error quoted is marginalized over the other two parameters, and
   are scaled to a 20,000 $deg^2$ survey, such as LSST.} }
\label{tab:merr}
\end{table}

\begin{figure}[htp]
\centering \includegraphics[width=7 cm]{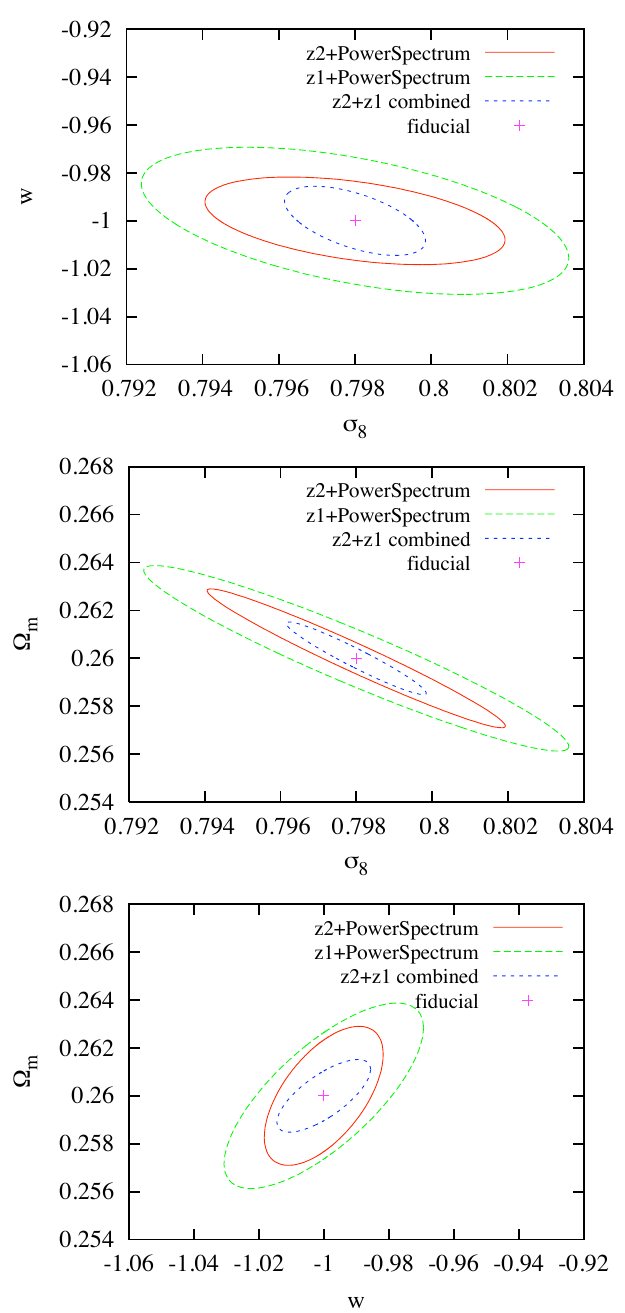}
\hfill
\caption[]{\textit{68\% percentile error ellipses in two-dimensional
    projections of the 3-dimensional parameter space of $\sigma_8$,
    $w$, and $\Omega_m$. In each panel, we show ``tomography'' results
    from noisy, unscaled maps for combining peak counts and power
    spectrum using either $z_s=2$, $z_s=1$, or their combination.  The
    constraints are scaled to a 20,000 $deg^2$ survey, such as LSST.}}
\label{fig:errorelipse}
\end{figure}

In all previous calculations, we have varied a single parameter,
holding all the other parameters fixed.  While this clarifies the raw
cosmological sensitivity of the peak counts, justified if CMB (or
other) observations can be used to determine the parameters with
negligibly small errors, one has to simultaneously vary all uncertain
parameter, and consider their degeneracies, to obtain realistic error
forecasts (even in the limiting case of no systematic errors).  While
numerical limitations preclude us from exploring the full cosmological
parameter space, we here use a Fisher matrix to obtain marginalized
errors when the three parameters $\sigma_8$, $w$, and $\Omega_m$ are
varied simultaneously.  Degeneracies between these parameters are
among the most important for both cluster
counts~(e.g. ref.~\cite{HMH01}) and for shear power
spectra~(e.g. ref.~\cite{DCPFCSD}).  We compute the marginalized
errors from Eq.~(\ref{eq:Fisher1}).  We use the finite difference
between the fiducial model and the low(high)-$\sigma_8$, low(high)-$w$
and low(high)-$\Omega_m$ models to estimate the backward(forward)
derivatives with respect to these parameters. The average of backward
and forward derivatives is used to calculate the Fisher matrix.

As mentioned above, we use 15 bins for the peak counts.  The simple
intuitive ad-hoc optimization of the bin boundaries, based on avoiding
crossings in a single pair of cosmologies, which was used in the case
of a single parameter, cannot be generalized in a straightforward way
to the multi-parameter case. Indeed, we have found that when we use
five bins, the results become sensitive to the choice of the cosmology
pair over which the bin boundaries are optimized.  Therefore we use
the simpler (and unambiguous) scheme of equal-count bins; as shown in
the previous section, the accuracy in this case convergences for
$\gsim 15$ bins.  We emphasize that whenever there are significant
degeneracies between parameters, the numerical accuracy requirements
on the individual elements of the Fisher matrix become more stringent.
on the individual elements of the Fisher matrix To validate our
results, we have checked that our marginalized errors do converge when
we use $\gsim15$ bins (see Fig.~\ref{fig:merr_binnum} above).

We found that, in addition to the binning, the marginalized errors for
the peak counts, with a single source galaxy redshift, are sensitive
to the direction of taking the finite-difference derivative (backward
or forward).  If we take any one of backward, forward and averaged
derivatives for any of the parameters: $\sigma_8$, $w$, $\Omega_m$,
among the marginalized errors of these 27 combinations, the
marginalized errors for the three parameters vary by about
$20\%-25\%$.  On the other hand, the results become more stable when
we combine the peak counts with the power spectrum (reducing the
variations to $10\% - 15\%$; to be consistent, we also use 15 bins for
the power spectrum for computing marginalized errors). This behavior
is consistent with the presence of strong degeneracies between
parameters, which are broken when Fisher matrices corresponding to two
or more observables are added (as shown for the combination of cluster
counts and power spectra~\cite{Wiley+04,FH07}).

In Table~\ref{tab:merr}, we show the results from combining peak
counts and power spectrum in noisy, unscaled maps, using either
$z_s=2$, $z_s=1$, or their combination.  All the numbers in this table
are scaled to the solid angle of $20,000~{\rm deg^2}$, representing an
all-sky survey such as LSST.  We simply divide our results from the
$12~{\rm deg^2}$ maps by a factor $\sqrt{20,000/12}$ (see discussion
in Paper I for this simple ``extrapolation'').  Comparing the
individual peak-count (1st row) and power spectrum (4th row) errors at
$z_s=2$ in the top half of the table with their combination (5th row),
we see that the individual errors are roughly similar, considering the
$20\%-25\%$ variation in the peak counts, whereas the combination
improves on either by a factor of $\approx$two.  As the table shows,
combining the peak counts and power spectrum at $z_s=2$ yield 68\%
constraints as tight as $(\Delta \sigma_8,\Delta\Omega_m,\Delta
w)=(0.0026, 0.0012, 0.019)$.  Combining the peak counts and power
spectrum at $z_s=1$ gives a constraints worse than at $z_s=2$, with
$\Delta \sigma_8$ and $\Delta\Omega_m$ larger by $\sim 30\%$ and
$\Delta w$ larger by $\sim 60\%$. This agrees with the previous
$\Delta\chi^2$ results that $z_s=2$ generally yields a better
sensitivity than $z_s=1$. We found that the marginalized errors from
two redshifts together decrease significantly compared to the
marginalized errors from $z_s=2$ alone. For the combination of the
peak counts and the power spectrum, we found $\Delta\sigma_8$ and
$\Delta\Omega_m$ are $\sim 50\%$ of the corresponding errors from
$z_s=2$, $\Delta w$ is $\sim 75\%$ of the error from $z_s=2$.  The
large decrease of marginalized error from tomography is different from
the results of $ \Delta\chi^2$ of peak counts: tomography does $not$
significantly improve the sensitivity due to the correlation between
two redshifts. This difference between $\Delta\chi^2$ (showing only
raw cosmological sensitivity) and marginalized errors (showing also
the degeneracy between cosmological parameters) clearly shows that the
change in peak counts induced by each parameter has a different
redshift-dependence. To be specific, the w-induced changes, as a
function of redshift, cannot be degenerate with, for example, the
Omega-induced changes (even if they can be very degenerate at a single
redshift, blowing up the marginalized errors).  Overall, these
constraints are comparable to those expected from other forthcoming
cosmology probes (see Paper I for a discussion), although a fair
comparison would involve replicating the cosmological parameter set,
and other assumptions made elsewhere, which is beyond the scope of
this study.

Our results are also shown graphically in Fig.~\ref{fig:errorelipse},
which show the 68\% joint two-parameter constraints (i.e.,
corresponding to $\Delta\chi^2=2.3$, and marginalized over the third
parameter).

\section{Summary and Conclusions}
\label{sec:conclusions}

In this paper, we used ray-tracing simulations, and a halo finder, to
study the halo contributions to peaks present in convergence maps,
expected in large forthcoming weak lensing surveys of the sky.  This
allowed us to understand the origin of relatively low-amplitude
``$0.5-1.5 \sigma$'', or ``medium'' peaks, and their sensitivity to
cosmology.  Our motivation to focus on these peaks is that they have
been shown to drive the overall cosmology-sensitivity of the peak
counts.  Given that weak lensing by large scale structure is among the
most promising cosmological datasets, expected to be available in the
near future, the cosmological information content of these robustly
measurable features must be understood.

We have found that unlike high peaks, which are typically produced and
dominated by a single collapsed halo, the medium peaks are primarily
caused by random noise. However, these medium peaks receive an
important contribution from a projection of multiple (typically, 4-8)
halos along the line of sight, which makes their number counts
sensitive to cosmological parameters.  We have shown that for source
galaxies at high redshift ($z_s=2$) the presence of noise {\em boosts}
the distinguishing power from peak counts -- a counter-intuitive
result that we have clarified analytically.

Our most important results are that the distribution of the medium
peaks differ from similar-height peaks in a pure Gaussian random field
(GRF).  We have shown, explicitly, that the peaks contain cosmological
information that differs from that in a GRF, and is non-degenerate
with the power spectrum of the convergence.  We have taken the first
steps toward more realistic error forecasts, by obtaining the
marginalized errors in the three--dimensional parameter space of
$\sigma_8$, $w$, and $\Omega_m$.  The results suggest that peak counts
will play a significant role in tightening cosmological constraints
from forthcoming large-solid-angle weak lensing surveys.  On the other
hand, we have not addressed here the long list of systematic errors
that will ultimately limit the utility of convergence peaks.  Given
that the sensitivity relies heavily on peaks whose height is close to
that of the expected noise, these issues will be especially important
to address in future work.  More generally, our results should motivate
investigations to extract yet more cosmological information from
nonlinear weak lensing features; Minkowski functionals appear to be a
promising possibility \cite{MFs}.

\begin{acknowledgments}
  We thank Lam Hui for helpful discussions. JMK would like to thank
  Kevin Huffenberger for useful discussions about statistics and the
  Fisher matrix. This work was supported in part by the NSF grant
  AST-05-07161, the U.S. Department of Energy under contract
  No. DE-AC02-98CH10886, the Initiatives in Science and Engineering
  (ISE) program at Columbia University, and the Pol\'anyi Program of
  the Hungarian National Office for Research and Technology
  (NKTH). JMK also received support from the University of Miami and
  from JPL subcontract 1363745.  This research utilized resources at
  the New York Center for Computational Sciences, a joint venture of
  Stony Brook University and Brookhaven National Laboratory located at
  Brookhaven National Laboratory which is supported by the
  U.S. Department of Energy under Contract No. DE-AC02-98CH10886 and
  by the State of New York. The simulations and WL maps were created
  on the IBM Blue Gene/L and /P New York Blue. The analysis was done
  on the LSST Linux cluster at BNL.
\end{acknowledgments}

\bibliographystyle{physrev}

\end{document}